%% file: 0689.tex
\documentclass[letterpaper]{aa520}
\usepackage{graphicx}
\begin{document}
%\thesaurus{}

\title{Modelling the spectral energy distribution of galaxies.}
\subtitle{III. Attenuation of stellar light in spiral galaxies}
\author{Richard. J. Tuffs\inst{1}, Cristina C. Popescu\inst{1,2}, Heinrich,
J. V\"olk\inst{1}, Nikolaos D. Kylafis\inst{3,4}, Michael A. Dopita\inst{5}}

\offprints{Richard.Tuffs@mpi-hd.mpg.de}

\institute{Max-Planck-Institut f\"ur Kernphysik,
           Saupfercheckweg 1, D-69117 Heidelberg\\
           \email{Richard.Tuffs@mpi-hd.mpg.de}\\ 
           \email{Cristina.Popescu@mpi-hd.mpg.de}
           \and
           Research Associate, The Astronomical Institute of the Romanian
           Academy, Str. Cu\c titul de Argint 5, Bucharest, Romania
           \and University of Crete, Physics Department, P.O. Box 2208, 710 03
           Heraklion, Crete, Greece
            \and Foundation for Research and Technology-Hellas, 
             71110 Heraklion, Crete, Greece
           \and
           Research School of Astronomy \& Astrophysics,
           The Australian National University, Cotter Road, Weston Creek ACT
           2611, Australia} 
            
\date{Received; accepted}

\abstract{We present new calculations of the attenuation of stellar light from
spiral galaxies using geometries for stars and dust which can reproduce the 
entire spectral energy distribution from the ultraviolet (UV) to the
Far-infrared (FIR)/submillimeter (submm) and can also 
account for the surface brightness distribution in both the
optical/Near-infrared (NIR) and FIR/submm. The
calculations are based on the model of Popescu et al. (2000), which 
incorporates a dustless stellar bulge, a disk of old stars with associated
diffuse dust, a thin disk of young stars with associated diffuse dust, and a
clumpy dust component associated with star-forming regions in the thin disk. 
The attenuations, which incorporate the effects of multiple anisotropic 
scattering, are derived separately for each stellar component, and
presented in the form of easily accessible polynomial fits as a function of 
inclination, for a grid in optical depth and wavelength. The wavelength 
range considered is
between 912\,${\AA}$ and 2.2\,${\mu}$m, sampled such that attenuation can be 
conveniently calculated both for the standard optical bands and for the bands 
covered by GALEX. 
The attenuation characteristics of the individual stellar
components show marked differences between each other. A
general formula is given for the calculation of composite attenuation, valid 
for any combination of the bulge-to-disk ratio and amount of
clumpiness. As an example, we show how the optical depth derived from the
variation of attenuation with inclination depends on 
the bulge-to-disk ratio. Finally, a recipe is given for a self-consistent 
determination of the optical depth from the ${\rm H\alpha}/{\rm H\beta}$ line 
ratio. 
}
\authorrunning{Tuffs et al. 2003}
\maketitle

\section{Introduction}

The measurement of star-formation rates and star-formation
histories of galaxies - and indeed of the universe as a whole -
requires a quantitative understanding of the effect of dust in attenuating  
the light from different stellar populations. Traditionally, the effect of 
dust has been
quantified by statistical analysis of the variation of optical surface
brightness with inclination (see Byun et al. 1994 and Calzetti 2001 for 
reviews). These studies often reached conflicting conclusions about the optical
thickness of galactic disks, and the need to solve this puzzle prompted the 
development of models for the propagation of light in spiral disks using 
radiation transfer calculations.

The first work of this type was that of Kylafis \& Bahcall (1987), who 
modelled the large scale distribution of stellar emissivity and dust of the 
edge-on galaxy NGC~891 using a finite disk and incorporating anisotropic and
multiple scattering. Using the same method, 
Byun et al. (1994) produced simulations of disk galaxies of 
various morphologies and optical thicknesses, while Xilouris et al. 
(1997, 1998, 1999) applied the technique to fit the optical/Near-infrared 
(NIR) appearance of further edge-on galaxies. Another approach was taken by
Witt et al. (1992), who studied the transfer of radiation
within a variety of spherical geometries, also incorporating anisotropic and
multiple scattering. This work was extended by Witt \& Gordon (1996, 2000) 
to include random distributions of dusty clumps within or around a smooth 
distribution of stars and further extended by Kuchinski et al. (1998) to 
model exponential disks and bulges of observed spiral galaxies (see also an
application of this technique to extremely red galaxies by Pierini et
al. 2004). A physical approach to clumpiness in terms of its relation to
interstellar turbulence and its effect on attenuation has been recently
presented by Fischera et al. (2004).

Specific studies of the attenuation of the spatially integrated light from
spiral galaxies have been made by Bianchi et al. (1996) and by Baes \& Dejonghe
(2001). Both works describe the influence of scattering and geometry on the
derived attenuation as a function of inclination, optical depth and
wavelength. Building on the work of Bianchi et al. (1996), Ferrara et
al. (1999) presented a grid of attenuation values corresponding to a range of
model galaxies spanning different bulge-to-disk ratios, opacities, relative
star/dust geometries and dust type (Milky Way and Small Magellanic Cloud type).

More powerful constraints on the geometrical distributions of stars and dust 
can be obtained by a joint consideration of the direct starlight, emitted in
the ultraviolet (UV)/optical/NIR, and of the starlight which is reradiated in 
the Far-infrared (FIR)/submillimeter (submm)\footnote{Based on recent ISO 
observations of a complete optically selected sample of normal late-type 
galaxies (Tuffs et al. 2002a,b), the percentage of stellar light reradiated 
by dust was found to account for $\sim30\%$ of the bolometric luminosity 
(Popescu \& Tuffs 2002a).}. In 
particular the study of Popescu et al. (2000) showed that, in addition to the
single exponential diffuse dust disk that was used to model the
optical/NIR emission, two further sources of 
opacity are needed to account for the observed amplitude and colour of the 
FIR/submm emission. Firstly, a clumpy and strongly heated distribution of 
dust spatially correlated with the UV-emitting young stellar population is
required to account for the FIR colours. This clumpy distribution of dust 
can most naturally be associated with the opaque parent molecular clouds of 
massive stars within the star-forming regions. It is also an integral
element of the model of Silva et al. (1998) for calculating the 
UV-submm spectral energy distribution (SED) of galaxies, and 
is not to be confused
with the randomly distributed clumps introduced by Witt \& Gordon (1996).
The second additional source of opacity is diffuse dust associated with the
spiral arms, which was approximated by a second exponential 
dust disk having the same spatial distribution as the UV-emitting stellar disk.
This component was needed to account for the amplitude of the submm emission.
Strong support for the reality of the elements of the model of Popescu et
al. (2000) comes not only from the success of the model to fit the integrated
SED, but also from the good agreement between the observed
surface brightness distributions in the FIR and those predicted by the model 
(Popescu et al. 2004). Additional
support is provided by the ability to predict UV magnitudes, as shown in
this paper.

In the first paper of this series (Popescu et al. 2000; hereafter Paper~I) we 
presented the general
technique\footnote{A simplified version of this technique has been recently 
applied by Misiriotis et  al. (2004) to fit the FIR SEDs of bright IRAS 
galaxies.}  for modelling the optical/NIR-FIR/submm SED (see also Popescu \&
Tuffs 2002b) and applied it to the 
edge-on galaxy NGC~891. In the second paper (Misiriotis et al. 2001; hereafter
Paper~II) the same 
model was successfully applied to further edge-on spiral galaxies
(NGC~5907, NGC~4013, UGC~1082, and UGC~2048), confirming that
the features of the solutions for NGC~891 are more generally applicable.
In the present paper we adopt these features as the basis for calculating the 
attenuation of stellar light, taking into account the constraints on 
the geometry of stars and dust arising from our fits to the 
optical-submm SEDs. These calculations are made for a grid in disk
opacity and inclination. We also generalise the model so that it is 
applicable to giant spiral galaxies with different amounts of clumpiness and 
with different bulge-to-disk ratios (Hubble types).
We emphasise that the model, in the form presented here, cannot be expected 
to work for small low 
luminosity systems like dwarf galaxies, which have systematically different 
geometries. Furthermore, 
the composition of dust in dwarf galaxies may differ from that in giant
spirals, resulting in a different extinction curve from the one found for 
NGC~891 and the other galaxies modelled in Paper~II.

A companion paper will present a corresponding grid of calculations for the
FIR/submm emission, thus providing a library of solutions of the SEDs over the
entire UV-submm range. This paper is organised as follows: In Sect.~2 we
present the model used for the calculation of attenuation. Sect.~3 describes
the use of radiation transfer calculations to obtain the attenuation in the
diffuse disk, thin disk and bulge components. In Sect.~4 we discuss
and compare the attenuation characteristics of the disk, thin disk and bulge. 
A formula and a recipe to
calculate composite attenuation of the integrated emission from galaxies as a
function of the parameters inclination, central face-on B-band optical depth, 
clumpiness and bulge-to-disk ratio is given in Sect.~5. In the same section 
we also discuss how to choose values for these parameters and we give an 
example for the
calculation of attenuation for the case of NGC~891 seen at different viewing
angles. To illustrate the use of composite attenuation curves in the
interpretation of optical/NIR data we consider in Sect.~6 the effect of a
varying bulge-to-disk ratio on the inclination dependence of the apparent
optical emission from galaxies. In Sect.~7 we give our summary and conclusions.
Readers interested only in practical applications of the model are directed to
Sect.~3 and 5, in particular to Tables\,4-6 and Eqs. 6, 17 and 18.

\section{The model}

In this section we describe the characteristics of the model adopted for the 
calculation of attenuation. In terms of geometry, the model can be divided 
into a diffuse component and a clumpy component, the latter being associated 
with the star-forming regions. Direct evidence for the existence of these two 
geometrical components comes from recent observations of resolved nearby 
galaxies done with the ISOPHOT instrument (Lemke et al. 1996) on board the 
Infrared Space Observatory (ISO; Kessler et al. 1996). For example the maps of
M~31 (Haas et al. 1998) and of M~33 (Hippelein et al. 2003) clearly show a
diffuse disk of cold dust emission prominent on the 170\,${\mu}$m images
as well as warm emission from HII regions along the spiral arms, prominent
on the 60\,${\mu}$m images. Warm and cold emission components have also been
inferred from statistical studies of local universe galaxies observed with 
ISOPHOT, in particular from mapping observations covering the whole
optical disk (Stickel et al. 2000 for serendipitously detected spirals; 
Popescu et al. 2002 for Virgo cluster galaxies).

\subsection{The diffuse component}

The diffuse component
is comprised of a diffuse old stellar population and associated dust and a
diffuse young stellar population and associated dust. The diffuse old stellar
population has both a bulge and a disk component, whereas the
diffuse young stellar population resides only in a thin disk. {\it 
Throughout this
paper we will use the superscript ``disk'', ``bulge'' and ``tdisk''  for all
the quantities describing the disk, bulge and thin disk. }

The emissivity of
the old stellar population is described by an exponential disk and a de
Vaucouleurs bulge:
\begin{eqnarray}\nonumber
\eta(\lambda,R,z) = {\eta^{\rm disk}}(\lambda,0,0) 
\exp \left( - \frac{R}{{h^{\rm disk}_{\rm s}}} - 
\frac{|z|}{{z^{\rm disk}_{\rm s}}} \right)
% {L_b} \exp (-7.67\,B^{1/4})\,B^{-7/8}~
\end{eqnarray}
\begin{eqnarray}
~~~~~~~~~~+{\eta^{\rm bulge}(\lambda,0,0)} 
\exp (-7.67\,B^{1/4})\,B^{-7/8}, 
\end{eqnarray}
\begin{eqnarray}
B = \frac{\sqrt{R^2 + z^2\,({a/b})^2}}{{R_e}},
\end{eqnarray}
where $R$ and $z$ are the cylindrical coordinates, 
$\eta^{\rm disk}(\lambda,0,0)$ is the stellar emissivity at the centre of the 
disk,
$h^{\rm disk}_{\rm s}$, $z^{\rm disk}_{\rm s}$ are the scalelength and 
scaleheight of the disk, $\eta^{\rm bulge}(\lambda,0,0)$ is the 
stellar emissivity at the centre of the bulge, $R_e$ is the effective radius 
of the bulge, and $a$ and $b$ are the semi-major and semi-minor axes of the 
bulge. 

The dust associated with the old stellar population is also described by an
exponential disk:
\begin{eqnarray}
\kappa^{\rm disk}_{\rm ext}(\lambda,R,z) = {\kappa^{\rm disk}_{ext}
(\lambda,0,0)}\,\exp \left( - \frac{R}{h^{\rm disk}_{\rm d}}- 
\frac{|z|}{z^{\rm disk}_{\rm d}} \right),
\end{eqnarray}
where $\kappa^{\rm disk}_{\rm ext}(\lambda,0,0)$ is the extinction coefficient
 at the centre of the disk and $h^{\rm disk}_{\rm d}$ and 
$z^{\rm disk}_{\rm d}$ are the scalelength and scaleheight of the dust 
associated with the old stellar disk.

The assumption that the scaleheight of the
old stellar population and associated dust is independent of 
galactocentric radius is worthy of some comment.
For an isothermal 
disk, the scaleheight, $z$, is related to 
the velocity dispersion $\sqrt\langle v^2\rangle$, and the density of matter, 
$\rho$, (both at the mid-plane) by 
$z \propto (\langle v^{2}\rangle/\rho)^{1/2}$ (Kellman 1972).
One can attempt to use this equation to infer the variation of the scaleheight
of the dust with galactocentric radius by considering the gaseous layer in
which the dust might be presumed to be embedded. Since the velocity dispersion
 of this gas has a value of about 10\,km\,s$^{-1}$, and is observed to remain 
approximately constant with
increasing radius (van der Kruit \& Shostak 1982, Shostak \& van der Kruit 
1984, Kim et al. 1999, Sellwood \& Balbus 1999, Combes \& Becquaert 1997),
the above equation would imply that in an exponential disk in which the matter
distribution follows that of the stellar luminosity, the scaleheight of the 
gas should slowly increase with radius, with a characteristic scalelength 
twice that of the exponential stellar disk. However, when the
additional gravitational force of the gas and dark matter halo is taken into
account (Narayan \& Jog 2002), the increase of scaleheight with radius is
predicted to be reduced. Also, the dust-to-gas ratio could well be a
decreasing function of $z$, further reducing the radial dependence of the
dust scaleheight. For the case of the stars, it is known that the velocity
dispersion falls exponentially with galactocentric radius (Bottema 1993),
reducing any increase of scaleheight with radius.
Therefore in the current generation of models we 
have not attempted to make refinements to our assumption that the scaleheights 
of stars and dust are independent of galactocentric radius.

For NGC~891 and other edge-on galaxies, Xilouris et al. (1999) derived the 
geometrical parameters in Eqs. 1-3 by fitting resolved 
optical and NIR images with simulated images produced from radiative transfer
calculations. They found scaleheights for the old stellar population of 
several hundred parsec, a result which had already been known for the
Milky Way, and which can be physically attributed to the increase of the 
kinetic temperature of stellar populations on timescales of order Gyr due to 
encounters with molecular clouds and/or spiral density waves (Wielen 1977). 
Another result to
emerge from this work was that the old stellar populations have scaleheights 
larger than those of the associated dust.
The resulting parameters for NGC~891, derived independently at each optical/NIR
wavelength by Xilouris et al. (1999), were used in Paper~I for 
the modelling of the UV-submm SED of this galaxy and were found
to be consistent with the FIR morphology of NGC~891 (Popescu et al. 2004).
Therefore we adopt these
parameters in this work, taking their averages
over the optical/NIR range, since no trend with
wavelength is apparent. The exception is the scalelength of the old stellar
disk $h^{\rm disk}_{\rm s}$ which decreases with increasing wavelength.
The averaged values describing the distribution of the old stellar population 
and associated dust are given in Table~1, where the length parameters have 
been normalised to the value of $h^{\rm disk}_{\rm s}$ in the B band. 
The adopted values for $h^{\rm disk}_{\rm s}$ (again normalised to the value 
of $h^{\rm disk}_{\rm s}$ in the B band) are given in Table~2, where the
I band value has been interpolated from the J and K band values.
We should stress that the normalised parameter values from Table~1 are
  close to the average values found by Xilouris et al. (1999) for his sample of
  edge-on galaxies. 

\begin{table} 
\caption{The parameters of the model. All length parameters are normalised to 
the B-band scalelength of the disk.}
\begin{tabular}{ll}
\hline 
$z^{\rm disk}_{\rm s}$ & 0.074 \\  
$h^{\rm disk}_{\rm d}$ & 1.406 \\  
$z^{\rm disk}_{\rm d}$ & 0.048 \\   
\hline
$h^{\rm tdisk}_{\rm s}$& 1.000 \\
$z^{\rm tdisk}_{\rm s}$& 0.016 \\
$h^{\rm tdisk}_{\rm d}$& 1.000 \\
$z^{\rm tdisk}_{\rm d}$& 0.016 \\
\hline
$R_e$                  & 0.229 \\
$b/a$                  & 0.6 \\
\hline
$\frac{\displaystyle \tau^{\rm f, disk}_{\rm B}}
{\displaystyle \tau^{\rm f, tdisk}_{\rm B}}$&  0.387\\
\hline 
\end{tabular}
\end{table}

\begin{table} 
\caption{Wavelength dependence of the scalelength of the disk normalised to 
its value in the B band.}
\begin{tabular}{lcccccccc}
\hline
                       &   UV   &  B   &  V   &  I   &  J   &  K  \\ 
\hline 
$h^{\rm disk}_{\rm s}$ &    -    & 1.000 & 0.966 & 0.869 & 0.776& 0.683\\
\hline
\end{tabular}
\end{table}

The emissivity of the young stellar population is also specified by an
exponential disk: 
\begin{eqnarray}
\eta^{\rm tdisk}(\lambda,R,z) = \eta^{\rm tdisk}(\lambda,0,0) \exp 
\left( - \frac{R}{h^{\rm tdisk}_{\rm s}} - \frac{|z|}{z^{\rm tdisk}_{\rm s}} 
\right)
\end{eqnarray}
where $\eta^{\rm tdisk}(\lambda,0,0)$ is the stellar emissivity at the centre 
of the thin disk and $h^{\rm tdisk}_{s}$ and $z^{\rm tdisk}_{s}$ are the 
scalelength and scaleheight of the thin disk.  

The dust associated with the young stellar population is again specified by an
exponential disk:
\begin{eqnarray}
\kappa^{\rm tdisk}_{ext}(\lambda,R,z) = 
\kappa^{\rm tdisk}_{ext}(\lambda,0,0)\,\exp 
\left( - \frac{R}{h^{\rm tdisk}_{\rm d}}- \frac{|z|}{z^{\rm tdisk}_{\rm d}} 
\right)
\end{eqnarray}
where $\kappa^{\rm tdisk}_{ext}(\lambda,0,0)$ is the extinction coefficient at
the centre of the thin disk and $h^{\rm tdisk}_{d}$ and $z^{\rm tdisk}_{d}$ are
the scalelength and scaleheight of the dust associated with the young stellar
disk. To minimise the number of free parameters, we fix the ratio
$\kappa^{\rm disk}_{ext}(\lambda,0,0)/\kappa^{\rm tdisk}_{ext}(\lambda,0,0)$ to
the value found for NGC~891 in Paper~I, and express it in terms 
of the ratio $\tau^{\rm f, disk}_{\rm B}/\tau^{\rm f, tdisk}_{\rm B}$ of 
central face-on optical depths in the B band (see Table~1). 
This is also close to what was found for NGC~5907 (Misiriotis et al. 2001).

The young stellar population is known to have smaller scale heights than the 
older stellar population and associated dust, and to be seen towards more 
optically thick lines of sight. Typically this population cannot be 
constrained from UV images, and therefore its scaleheight 
$z^{\rm tdisk}_{\rm s}$ was simply
fixed to be 90\,pc (close to that of the Milky Way; Mihalas \& Binney 1981)
and its scalelength $h^{\rm tdisk}_{\rm s}$ was equated to the scalelength of 
the old stellar population in the B band (see Table~1). The dust associated
with the young stellar population was fixed to have the same scalelength and
scaleheight as for the young stellar disk, namely  
$h^{\rm tdisk}_{\rm d}$=$h^{\rm tdisk}_{\rm s}$ and 
$z^{\rm tdisk}_{\rm d}$=$z^{\rm tdisk}_{\rm s}$ (see Table~1). The reason for
this choice is that our thin disk of dust was introduced to mimic the diffuse
component of dust which pervades the spiral arms, and which occupies the same 
volume as that occupied by the young stars. This choice is also physically
plausible, since the star-formation rate is closely connected to the gas
surface density in the spiral arms, and this gas bears the grains which caused
the obscuration. In principle, we might expect that the metallicity
gradient of the gas within the galaxy would decrease the radial scalelength of
the dust ($h^{\rm tdisk}_{\rm d}$) below that of the gas. However, the ratio 
of the gas to stellar surface densities increases with galactocentric radius,
and so tends to cancel any variation in the ratio of the radial scalelength 
of the dust with respect to that of the young stars.

\subsection{The clumpy component}

By their very nature, star-forming regions harbour optically thick clouds which
are the birth places of massive stars. There is therefore a certain probability
that radiation from massive stars will be intercepted and absorbed by their 
parent clouds. This process is accounted for by a clumpiness factor $F$ which 
is defined as the total fraction of UV light which is locally absorbed in the
star-forming regions where the stars were born. Astrophysically this process 
arises
because at any particular epoch some fraction of the massive stars have not had
time to escape the vicinity of their parent molecular clouds. Thus, $F$ is
related to the ratio between the distance a star travels in its lifetime due to
it's random velocity and the typical dimensions of star-forming complexes.
To conclude, in our formulation the clumpy distribution of dust is 
  associated
  with the opaque parent molecular clouds of massive stars, and is not to be
  confused with the randomly distributed clumps (of dust unrelated to the
  stellar sources).

The attenuation by the clumpy component has a different behaviour than that of
the diffuse component. One difference is that the attenuation 
by the clumps is independent of the inclination of the galaxy. Another 
difference is that the
wavelength dependence is not determined by the optical properties of the
grains (because the clouds are so opaque that they block the same proportion of
light from a given star at a given time at each wavelength), but instead 
arises because stars of different masses survive for
different times, such that lower mass and redder stars can escape further from
the star-forming complexes in their lifetimes. A proper treatment of 
the clumpiness factor is important since the clumpiness will change the shape
and the inclination dependence of the UV attenuation curves of star-forming 
galaxies.

\subsection{The dust model}

The dust model corresponds to the 
graphite/silicate mix of Laor \& Draine (1993) and to the $a^{-3.5}$ grain size
distribution of Mathis, Rumple \& Nordsieck (1977) and was also used in the 
calculation of the optical-submm emission in Paper~I.
The wavelength dependence of the extinction coefficient $\kappa_{\rm ext}$, 
albedo and the anisotropy factor $g$  for this model is given in Table~3.  This model is consistent with a Milky Way extinction curve and, as 
already noted, with the extinction curves found for the galaxies modelled in
Paper~II.

\begin{table}
\caption{The wavelength dependence of the normalised extinction coefficient 
$\kappa_{\rm ext}^{\star}$ (normalised to its value in the B band), albedo 
and the anisotropy factor $g$.}
\begin{tabular}{llll}
\hline
$\lambda$$^{\dagger}$ & $\kappa_{\rm ext}^{\star}$ & albedo & $g$\\
 $\AA$    &                             &       &     \\
\hline
  912 & 4.82 & 0.28 & 0.57 \\
 1350 & 2.57 & 0.38 & 0.62 \\
 1650 & 2.04 & 0.43 & 0.61 \\
 2000 & 2.44 & 0.43 & 0.53 \\
 2200 & 2.60 & 0.43 & 0.49 \\
 2500 & 1.99 & 0.52 & 0.48 \\
 2800 & 1.66 & 0.54 & 0.49 \\
 4430 & 1.00 & 0.53 & 0.50 \\
 5640 & 0.74 & 0.52 & 0.45 \\
 8090 & 0.44 & 0.49 & 0.36 \\
12590 & 0.20 & 0.37 & 0.15 \\
22000 & 0.06 & 0.16 & 0.04 \\
\hline
\end{tabular}

$^{\dagger}$The wavelengths are those for which the attenuation is calculated
 (see Sect.~3).
\end{table}

\subsection{The free parameters of the model}

The optical-submm SED model presented in Paper~I has only three free 
parameters, since, as described before, the
geometry is constrained by optical/NIR images. The three free parameters are:
$SFR$, mass of dust in the thin disk, and the clumpiness factor $F$. For
the specific application of calculating attenuation in the UV to NIR spectral
range, $SFR$ is not needed, since
extinction does not depend on the strength of sources. Also, because here we
have fixed the ratio between the $\tau_{\rm B}^{\rm f, disk}$ and 
$\tau_{\rm B}^{\rm f, tdisk}$ (see Table~1), and because the dust model is
also fixed, the mass of dust in the thin disk is fully determined by the total 
central face-on optical depth in B band, $\tau_{\rm B}^{\rm f}=
\tau_{\rm B}^{\rm f, disk}+\tau_{\rm B}^{\rm f, tdisk}$. The third free
parameter for the calculation of the whole SED - the clumpiness factor $F$ - 
remains a free parameter for the calculation of attenuation. In addition,
attenuation also depends on the inclination $i$. In summary, we need three
parameters to fully determine the attenuation in a galaxy: $\tau_{\rm B}^{\rm
f}$, $F$ and $i$. A fourth free parameter - the bulge-to-disk ratio - is
introduced in Sect.~5 to account for the different morphologies encountered in
the Hubble sequence of spiral galaxies. 

\begin{figure*}[htb]
\includegraphics[scale=0.75]{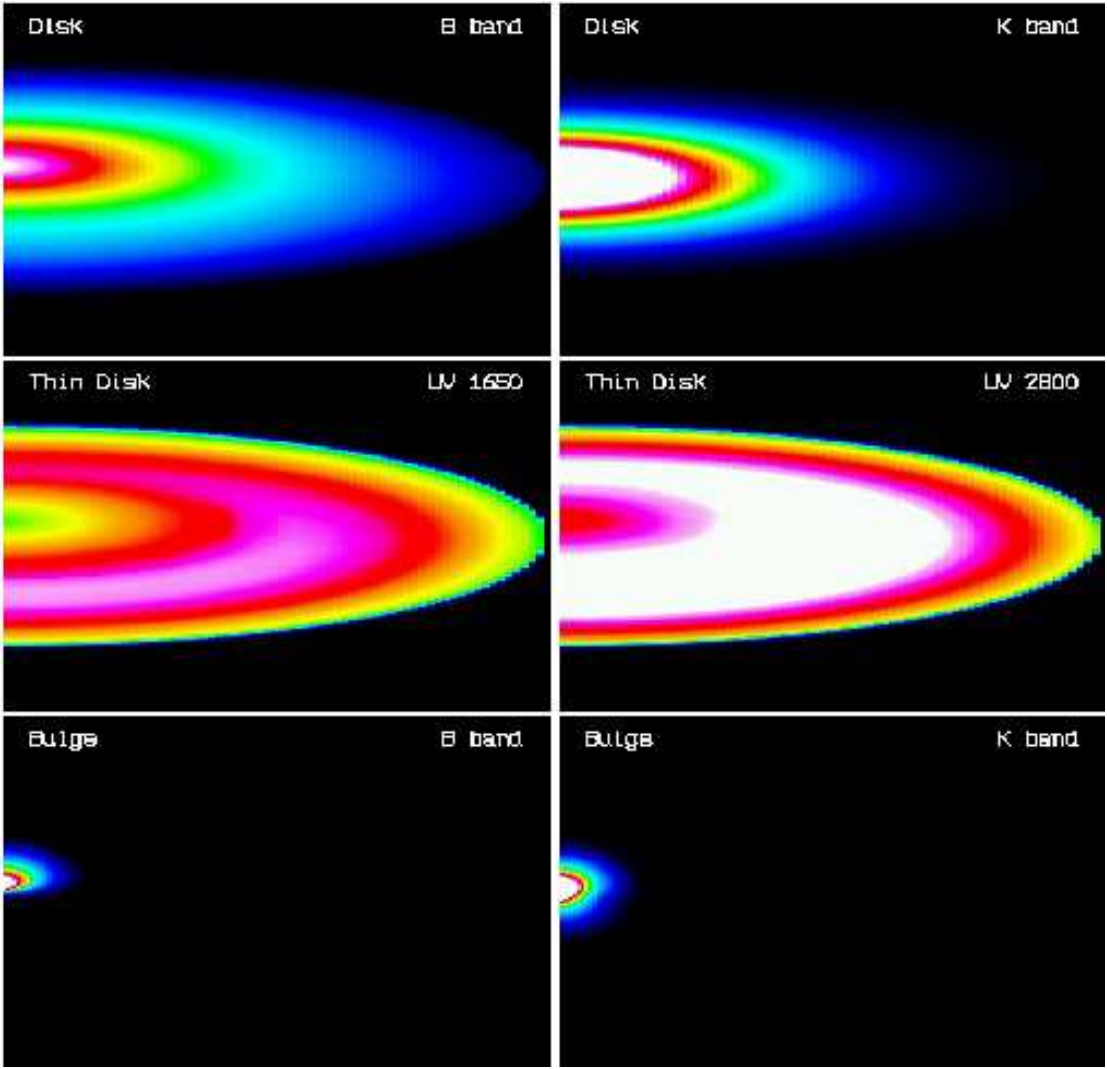}
\caption{Simulated dust-attenuated images of the disk, thin disk and bulge 
seen at 78 degrees inclination, for a total central face-on optical 
depth in the B band $\tau^{\rm f}_{\rm B}=4$. Only the right hand side of each
image is displayed, with the centre of the galaxy lying half way along
the left hand border of each panel. The dust disks which produce the
attenuation are inclined such that they are closer to the observer
on the lower half of each image. All images have been displayed 
with the same spatial scale and have the same colour table. For each pair of 
images, pixels with the same intensity are displayed with the same colour. 
Each image was scaled such that the brightest pixel on the corresponding 
unattenuated image has the value 1.0. In this way one can
see the effect of increased opacity with decreasing wavelength, by
comparing the left hand panel with the right hand panel, for each geometrical
component.  The top panels show the disk in the B band (left) and K band
(right). The smaller extent of the disk as seen in the K band is due to the 
smaller scalelength of the stellar emissivity in the K band compared to that in
the B band. In both images the maximum brightness comes from the central
regions, though at B band the area of bright emission (white colour) is smaller
than that in the K band, due to the increased opacity. The middle panels show
the thin disk at 1650\,$\AA$ (left) and 2800\,$\AA$ (right). In both images the
maximum brightness is no longer in the central region, but at an intermediate
radius, leaving a central ``hole'', whose area (red colour) is larger at the 
shorter UV wavelength. The asymmetry around the horizontal axis is caused by 
the stronger forward scattering which brightens the near side of the thin disk
compared to the far side. The bottom panels show the bulge in the B (left)
and K band (right). In the K band bulge emission from both above and below 
the dust disks is seen, though the emission from below is somewhat attenuated.
In the B band almost all the emission from below the dust disk is obscured.}
\end{figure*} %Fig 1

\section{Calculation of attenuation in the diffuse component}

This section describes the use of radiation transfer calculations to obtain the
attenuation in the diffuse component. No radiation transfer calculations 
are needed for the clumpy component, which is handled analytically 
(see Sect.~5).

\begin{figure*}[htb]
\includegraphics[scale=0.8]{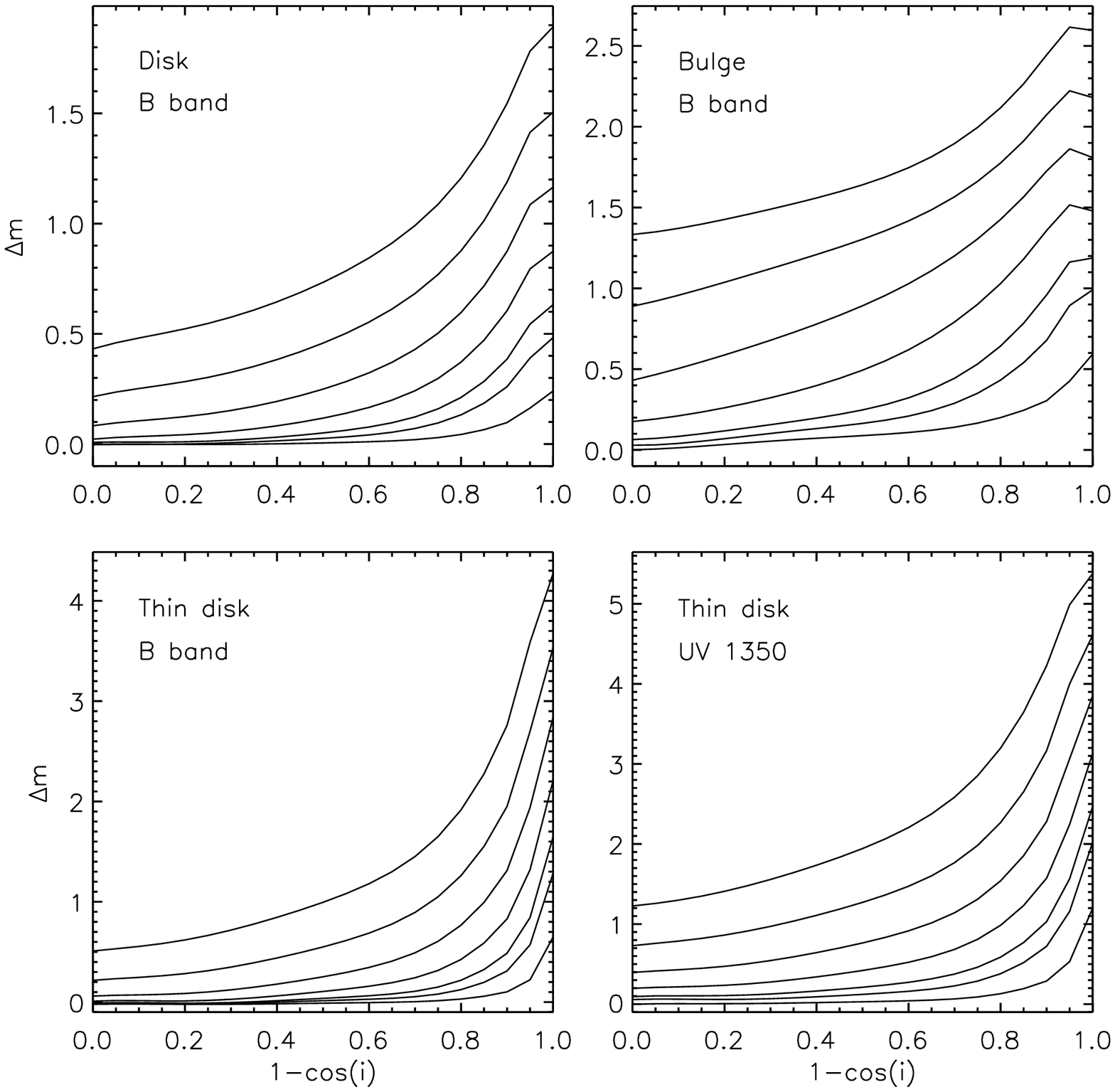}
\caption{Examples of the dependence of attenuation ($\Delta m$) on 
inclination ($i$) for the main
geometrical components of our model: disk (top left), bulge(top right), and thin
disk (bottom left and right). The examples are plotted for the B band for the
disk and bulge, and both for the B band and UV 1350$\AA$ for the thin disk. In
each panel we plotted (from top to bottom) 7 attenuation curves, 
corresponding to $\tau^{\rm f}_{\rm B}$: 8, 4, 2, 1, 0.5, 0.3, and 0.1. The 
face-on orientation corresponds to $1-\cos(i) = 0.0$ and the edge-on 
orientation corresponds to  $1-\cos(i) = 1.0$.}
\end{figure*} %Fig 2

The basic approach used here is to calculate the attenuation separately for 
the three
diffuse geometrical components of our model: the disk, the thin disk and 
the bulge. In each case the same fixed geometry of the diffuse dust is 
adopted, 
which is the superposition of the dust in the disk and the dust in the thin 
disk. In other words we derive the attenuation of the old stellar population
in the disk as seen through the dust in the disk and thin disk; we derive the
attenuation of the young stellar population in the thin disk as seen though the
same dust in the disk and thin disk; and we derive the attenuation of the
bulge, also viewed through the dust in the disk and thin disk.

The calculations were performed for combinations of the two parameters
affecting the diffuse component,  $\tau_{\rm B}^{\rm f}$ and inclination
$i$. For the sampling in $\tau_{\rm B}^{\rm f}$ we chose the set of values: 
${0.1,0.3,0.5,1.0,2.0,4.0,8.0}$ which range from extremely optically thin to
moderately optically thick cases.\footnote{We should remark that, since  
$\tau_{\rm B}^{\rm f}$ is the opacity through the centre of the galaxy, where
most of the dust is concentrated, a choice of 1.0 for $\tau_{\rm B}^{\rm f}$
actually represents an optically thin galaxy over almost all of its area. Even
for $\tau_{\rm B}^{\rm f}=4$, which is close to that found for NGC~891, 
less than half of the total bolometric luminosity is absorbed by dust.}
For the sampling in inclination we chose $0 \leq \cos(i) \leq 1$, with 
$\Delta \cos(i) = 0.05$. Each calculation was performed at a different
wavelength, covering the whole UV/optical/NIR range, such that attenuation 
can be conveniently calculated for both standard optical bands and the bands
covered by GALEX. Our choice of UV wavelengths also samples the 2200\,$\AA$ 
feature. The calculations in
  the UV range were only performed for the thin disk, since this is the only
  component of stellar emissivity emitting in this spectral range. In total 
we used 12 wavelengths, listed in Table~3. 

Simulated images of the pure disk, thin disk and bulge were calculated 
for each combination of parameters using the radiative transfer code of 
Kylafis \& Bahcall (1987), which includes anisotropic multiple scattering. In 
total we produced 3234 images with a pixel size (equal to the resolution) of 
0.0066 of the B-band scalelength $h^{\rm disk}_{\rm s}$, sampled every 5 and 10
pixels in the inner and outer disk, respectively. The high resolution of the 
simulated images matches
the resolution of the optical images of NGC~891 used in the optimisation
procedure for the derivation of the disk/bulge geometry. This choice enables 
not
only high accuracy in the derivation of attenuation, but also means that the
resulting model images (after suitable interpolation) can be used as 
template images for comparison with
observed images of real galaxies. The simulated images for the disks extend 
out to a radius of 4.63 B-band scalelengths $h^{\rm disk}_{\rm s}$, which is 
equivalent to 3.31 dust scalelengths $h^{\rm disk}_{\rm d}$. The simulated 
images for the bulge extend out to 1.45 B-band scalelength 
$h^{\rm disk}_{\rm s}$. For both the disks and the bulge the extent of the
integration is a numerical limit. Examples of
simulated images for the disk, thin disk and bulge are given in Fig.~1.

We also produced the 
corresponding intrinsic images of the stellar emissivity (as would be 
observed in the absence of dust). The attenuation ${\Delta}m$ was then 
obtained by
subtracting the integrated magnitude of the dust affected images from the
integrated magnitude of the intrinsic images.

At this point we have obtained values of the attenuation for different
combinations of $i$ and $\tau^{\rm f}_{\rm B}$, at each wavelength, and 
independently for the disk, thin disk and bulge. To facilitate access to this
information and to allow attenuation to be calculated for any inclination, we
fit the attenuation curves (${\Delta}m$ vs $i$) with polynomial functions 
of the form:
\begin{eqnarray}
\Delta m = \left\{
  \begin{array}{lll}
   \displaystyle \sum_{{\rm j}=0,n} a_{\rm j}(1-\cos(i))^{\rm j} 
                                    & {\rm for} & 1-\cos(i) \leq 0.90\\
   b_0                              & {\rm for} & 1-\cos(i) = 0.95\\
   b_1                              & {\rm for} & 1-\cos(i) = 1.00\\
  \end{array}
  \right .
\end{eqnarray}
where $n=5$ for the disk and thin disk and $n=4$ for the bulge.
Values for the coefficients $a_j$ and for the constants $b_0$ and $b_1$ are
given in Tables~4, 5 and 6 for the disk, thin disk and bulge, respectively.
To derive the attenuation for any desired combination of $i$ and $\tau^{\rm
f}_{\rm B}$ one can first apply Eq.~6 to obtain the values of ${\Delta}m$
corresponding to the sampled $\tau^{\rm f}_{\rm B}$ and then
interpolate between these values. Subsequently one can also interpolate in
wavelength.
 
It should be noted that slight negative values for $\Delta m$ are 
obtained for some combinations of low inclination and low 
$\tau^{\rm f}_{\rm B}$. This mild amplification is due to the scattering of
light which removes photons travelling at high inclinations (in the plane of 
the disk) and sends them into directions with low inclinations, as also shown
by Baes \& Dejonghe (2001) for the case of isotropic scattering.

In Fig.~2 we show examples of the dependence of attenuation on inclination for
the three geometrical diffuse components, both in the optical and in the
UV. As expected, the increase in attenuation with increasing inclination is
stronger for larger $\tau^{\rm f}_{\rm B}$ than for lower 
$\tau^{\rm f}_{\rm B}$, irrespective of geometry or wavelength.

\section{Attenuation characteristics for the disk, thin disk and bulge}

\subsection{Attenuation of the disk}

In Fig. 3 (upper four panels) we show the wavelength dependence of the 
attenuation of the disk, from the B band to the K band. As expected, the 
overall level of the 
attenuation increases with increasing $\tau^{\rm f}_{\rm B}$. 
Another feature is the bunching of the
curves at low inclinations and low $\tau^{\rm f}_{\rm B}$, followed by an 
increase in the spacing of the curves when proceeding to higher inclinations 
and higher $\tau^{\rm f}_{\rm B}$. This can be 
explained as follows. At low inclination and $\tau^{\rm f}_{\rm B}$ most of 
the disk is optically thin and the attenuation will scale as
$-log(1-\tau)$, where $\tau$ is the line of sight optical depth. Thus, in 
this regime, changes in inclination or 
$\tau^{\rm f}_{\rm B}$ induce only small changes in apparent
magnitude. However, as we proceed to higher inclinations and 
$\tau^{\rm f}_{\rm B}$, the disk area which is optically thick will 
increase from the centre, spreading to the outside. In the inner optically 
thick part of the disk attenuation scales linearly with $\tau$ whereas in the 
outer optically thin parts the attenuation still scales logarithmically. Thus 
the increasing fraction of the disk area which is optically thick will have the
effect of inducing bigger and bigger changes in the apparent magnitudes. 

\begin{figure*}[hp]
\includegraphics[scale=0.76]{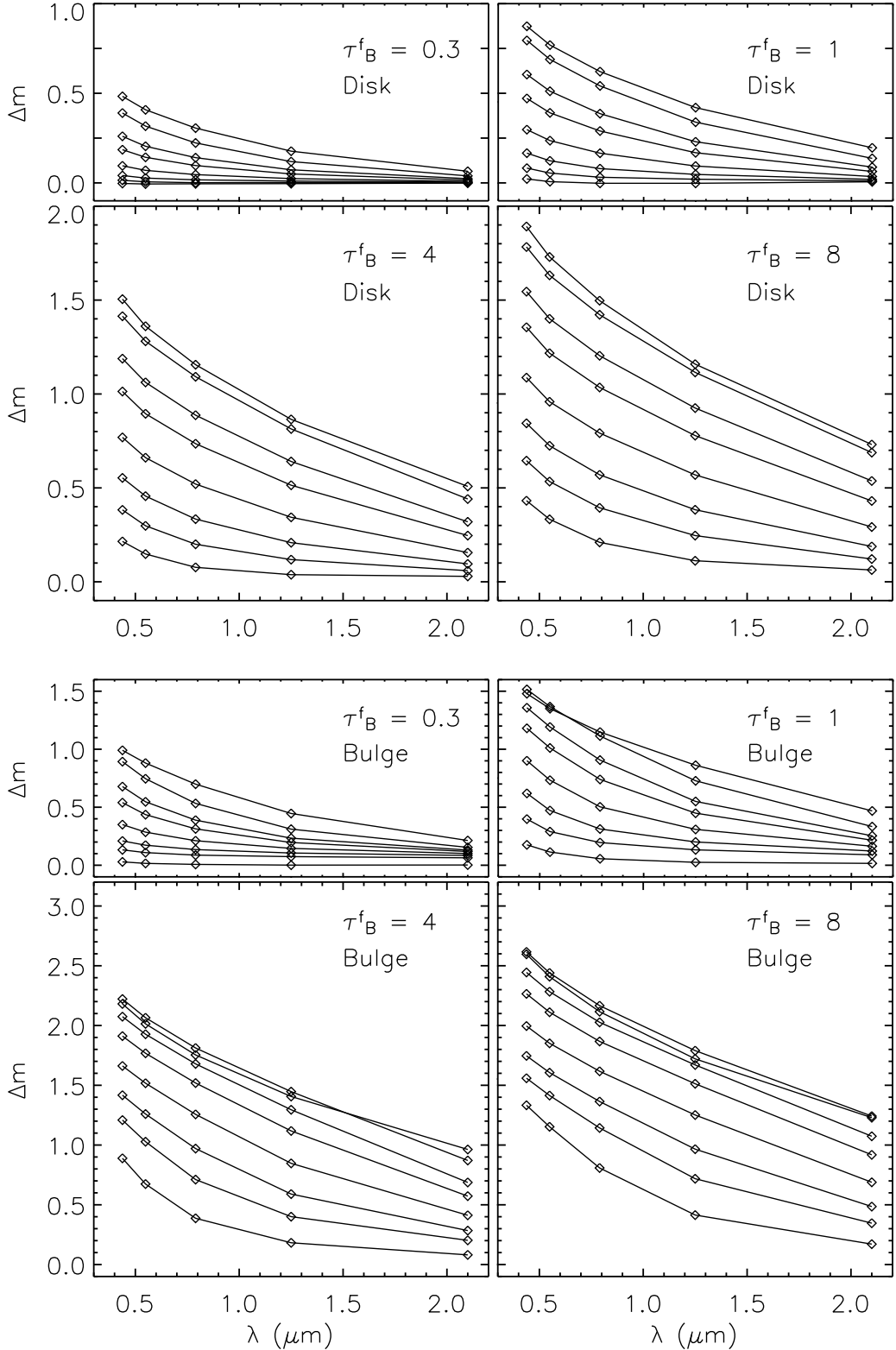}
\caption{The dependence of the attenuation ($\Delta m$) of the disk (upper four
  panels) and of the bulge (lower four panels) on 
optical/NIR wavelength ($\lambda$). In each case the panels display the 
attenuation curves for different $\tau^{\rm f}_{\rm B}$: 0.3, 1, 4 and 8.
In each panel the symbols represent attenuations calculated for 5 
optical/NIR wavelengths: B, V, I, J and K band. The solid lines represent the
linear interpolation between adjacent calculated points. 
 In each panel we
plotted (from top to bottom) 8 attenuation curves, corresponding to
inclinations $1-\cos(i)$: 1, 0.95, 0.90, 0.85, 0.75, 0.60, 0.40, 0.}
\end{figure*} %Fig. 3

The bunching of the curves of attenuation versus wavelength at low 
inclination and low $\tau^{\rm f}_{\rm B}$ can also be
appreciated from examination of the curves of attenuation versus inclination
shown in Fig.~2a (for the case of the B band). The bottom 
curves of Fig.~2a, corresponding to low $\tau^{\rm f}_{\rm B}$, are almost
flat over most of the inclination range, meaning that the increase in
inclination produces only small changes in attenuation. Similarly, the 
increase in the spacing of the curves of attenuation versus wavelength at 
high inclination and high $\tau^{\rm f}_{\rm B}$ from Fig.~3 (upper four
panels) is reflected by 
the upper curves of Fig.~2a (corresponding to high $\tau^{\rm f}_{\rm B}$),
which continuously steepen.

 A further feature of the disk attenuation is the smaller spacing of the 
curves between the edge-on geometry and the slightly inclined from edge-on 
geometry (the two top curves in each of the upper four panels of Fig.~3), an 
effect which 
becomes more pronounced at high $\tau^{\rm f}_{\rm B}$. This effect can also be
seen in Fig.~2a for the case of the B band, where the curves of attenuation 
versus inclination flatten when approaching the edge-on viewing angle.  
This is because the stellar disk has a higher scale height than the dust. In 
the edge-on view, 
the high z tail of the stellar population will be visible both above and 
below the plane. This will tend to boost the apparent luminosity of the disk 
compared with the slightly inclined disk, partially cancelling the dimming due
to the increased line-of-sight optical depth.

\subsection{Attenuation of the bulge}

Analogous to the case of the disk, we present the wavelength dependence of the
attenuation of the bulge in Fig~3 (lower four panels). The most interesting 
result is the larger overall attenuation of the bulge, as compared to that of 
the disk (Fig~3, upper four panels). This is because the bulge is more 
concentrated towards the 
optically thick part of the dust disks. For the same reason the curves also 
exhibit a strong dependence of attenuation on inclination. Especially for low 
$\tau^{\rm f}_{\rm B}$, this dependence is even more pronounced than for the
disks. At high $\tau^{\rm f}_{\rm B}$ though, the variation of attenuation with
inclination is limited by the larger vertical extent of the stellar population
of the bulge compared to the vertical extent of the dust. 

Another feature of the attenuation curves for the bulges is the crossing of 
the two uppermost curves, for 87 and 90 degrees (Fig.~3, lower four panels).
The intersection of the curves occurs at a critical wavelength
(which depends on $\tau^{\rm f}_{\rm B}$), shortwards of which the
highest attenuation is no longer for 90 degrees inclination, but instead at a 
slightly smaller inclination. This critical wavelength increases with 
increasing $\tau^{\rm f}_{\rm B}$: it lies between B and V bands for 
$\tau^{\rm f}_{\rm B}=1$ , between V and I bands for $\tau^{\rm f}_{\rm B}=2$,
 between I and J bands for $\tau^{\rm f}_{\rm B}=4$ and goes to K band 
for $\tau^{\rm f}_{\rm B}=8$.
The explanation for these features is as follows. If we consider that the 
dust disk is optically thick for all lines of sight intersecting the bulge, 
then the fraction of light
blanked out is simply proportional to the projected area of the disk onto the
bulge. Since the projected area of the disk is at a minimum for an inclination
of 90 degrees, moving away from the edge-on geometry will increase the
attenuation. If we progress to even lower inclinations the disk will start to
become optically thin for some lines of sight through the bulge, and then the
attenuation will start to decrease with decreasing inclination. This
non-monotonic progression of attenuation with inclination close to the edge-on
orientation is also illustrated in Fig.~2b (for the example of
the B band). There one can see that the maximum attenuation is no longer reached
at 90 degrees for $\tau^{\rm f}_{\rm B} \ge 1$. 

\begin{figure*}[hp]
\includegraphics[scale=0.76]{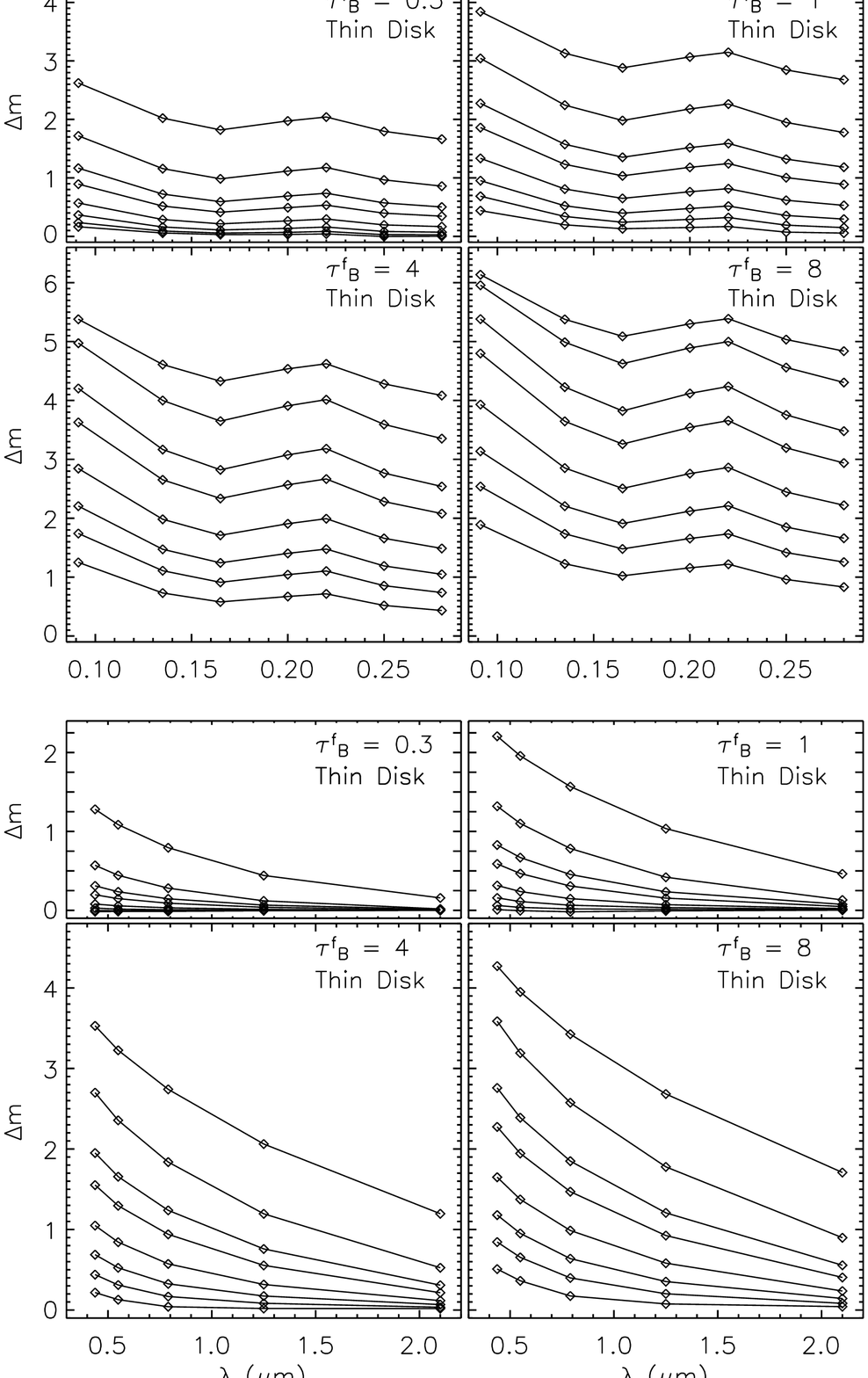}
\caption{The dependence of the attenuation ($\Delta m$) of the thin disk on 
UV wavelength (upper four panels) and on optical/NIR wavelength
(lower four panels). Curves are plotted as in Fig.~3, except that for the upper
four panels the symbols now represent attenuations calculated for 7 UV 
wavelengths: 912, 1350, 
1650, 2000, 2200, 2500 and 2800\,$\AA$.}
\end{figure*} %Fig. 4

\subsection{Attenuation of the thin disk}

In Fig.~4 (upper four panels) we show the dependence of the attenuation of 
the thin disk on UV
wavelength. In this wavelength range the opacity does not decrease 
monotonically with increasing wavelength, as it does in the optical regime, 
because
of the local maximum around 2200\,$\AA$ in the extinction efficiencies. This
causes the bump around 2200\,$\AA$ seen in the attenuation curves in Fig.~4
(upper four panels).

As in the case of the attenuation of the disk (Fig.~3, upper four panels), 
one can observe a
bunching of the curves for low inclinations and low $\tau^{\rm f}_{\rm B}$,
followed by an increase in the spacing between curves in proceeding to high 
inclinations and high $\tau^{\rm f}_{\rm B}$. In addition to the factors
contributing to this effect discussed in Sect. 4.1, scattering of light also
payes a role. Photons travelling at high inclinations (in the plane of the 
disk) are scattered into directions with low inclinations, such that they can 
escape the disk (as explained by Baes \& Dejonghe 2001 for the case of 
isotropic scattering). This factor also influences the attenuation in the 
disk, but is more important in the case of the thin disk, since here the 
stars have a stronger correlation with the dust, so that the contrast between 
the edge-on and the face-on optical depth is higher. 

\begin{figure*}[htb]
\includegraphics[scale=0.8]{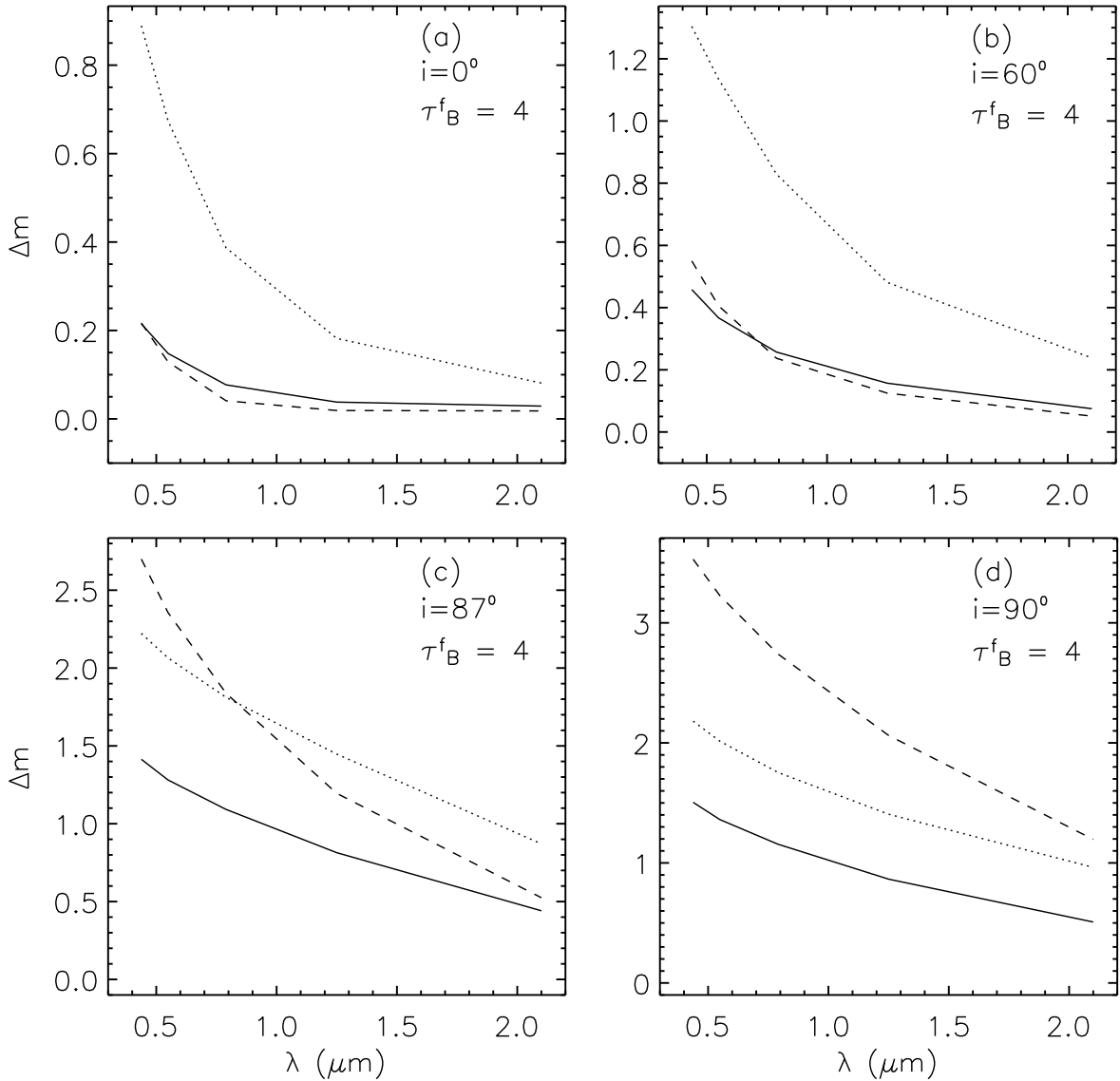}
\caption{Comparison between the attenuation curves of the disk, thin disk and
bulge. The 4 panels show the dependence of the attenuation on optical
wavelength, for 4 inclinations: 0, 60, 87 and 90 degrees, and for the same 
$\tau^{\rm f}_{\rm B}=4$. In each panel the attenuation for the disk, thin disk
and bulge are plotted with solid, dashed and dotted lines, respectively.}
\end{figure*} %Fig. 5

At high $\tau^{\rm f}_{\rm B}$, however, there is a small tendency for the 
curves for the edge-on and nearly edge-on inclinations to come together. This 
effect, which is most 
prominent for 912\,$\AA$ and $\tau^{\rm f}_{\rm B}=8$ (Fig.~4, upper four
panels), is primarily a saturation effect due to the fact that only a
  thin skin is seen close to edge-on.

In Fig.~4 (lower four panels) we show the dependence of the attenuation of 
the thin disk on optical
wavelength. At larger inclinations, the thin disk exhibits the largest
attenuation in the optical range, as compared with the disk and the bulge. This
is because of the strong spatial correlation between the stars and the
dust. For the same reason, the thin disk also exhibits the strongest dependence
on inclination. Despite the larger
attenuation, the contribution of this geometrical component to the overall
attenuation of the galaxy in the optical range is small and can be
neglected, since the thin disk emits mainly in the UV. However the attenuation
of the thin disk in the optical range is important for the calculation of
optical line emission arising from the thin disk.

\subsection{Intercomparison  of the attenuations of the disk, thin disk and 
bulge}

In the previous subsections we described and discussed the main characteristics
of the attenuation curves of the main geometrical components of our model,
namely of the disk, thin disk and the bulge. To make an intercomparison of the
curves for the different geometries we superpose in Fig.~5 attenuation curves 
for the three geometrical components, for the same inclination and 
$\tau^{\rm f}_{\rm B}$. The value of $\tau^{\rm f}_{\rm B}$ was chosen to 
be 4, which, for the geometry of our model, is close to what we found for
NGC~891.

At low and intermediate inclinations the bulge suffers a larger
  attenuation than that of the disk or thin disk (Fig.~5a,b). As already 
noted, this is because 
the bulge is more concentrated towards the optically thick part of the dust 
disks. In addition, and for the same reason, the figure also shows that the 
attenuation of the bulge has the steepest dependence on optical wavelength. 
It is also apparent from Fig.~5a,b that the attenuation curves for the disk 
and thin disk show quite similar behaviour to each other, indicating that at 
low and intermediate inclinations the curves are not sensitive to the 
differences in scale height and scale length between the two disks. There is
nevertheless a small difference between the curves for the disk and thin disk,
in the sense that the thin disk suffers slightly less attenuation, except for
the B band. This is because the disk has smaller scale lengths for the 
stellar population than the thin disk, except for the B band, where the
scale length is the same.

With increasing line of sight optical depth, which is the same as increasing 
inclination and/or decreasing wavelength, the attenuation curve for the thin 
disk starts to diverge from that of the disk (Fig~5c); it will increase until
it intersects with the attenuation curve for the bulge. This is because the
thin disk has the smallest scale height, the strongest spatial correlation
with the dust and the largest contrast between the face-on and edge-on
orientation, and so its attenuation depends most
strongly on inclination (or wavelength). So for edge-on orientation (Fig.~5d) 
the attenuation of the thin disk dominates that of the other geometrical 
components, reversing the situation for the face-on orientation (Fig.~5a), 
where the thin disk has the smallest attenuation.

\section{Composite attenuation curves for spiral galaxies}

Spiral galaxies have varying proportions of their stellar luminosities
emitted by the disk, thin disk and bulge and also varying amounts of
clumpiness.  For instance the bulge-to-disk ratio decreases in going 
from earlier to later spiral types (see for example Fig.~7 of Trujillo et al. 
2002). In the previous sections we have calculated attenuation curves 
quantifying the extinction of the individual stellar populations from the 
disk, thin disk and bulge due to the diffuse dust. These can be combined to 
obtain a composite attenuation curve
for the light illuminating the diffuse dust in a galaxy. If we also consider  
the photons that don't reach the diffuse dust because they are absorbed by the
clumpy dust, local to the star-forming regions, then we obtain the composite
attenuation curve for the total luminosity of the galaxy.

\subsection{A formula to calculate composite attenuations}

At a given wavelength $\lambda$, the total attenuation ${\Delta} m_\lambda$ 
in a galaxy is given by:
\begin{eqnarray}
{\Delta} m_\lambda & = & -2.5\log\frac{S_\lambda}{S^{\rm 0}_\lambda},
\end{eqnarray}
where $S^{\rm 0}_\lambda$ and $S_\lambda$ are the intrinsic and the apparent
flux densities, respectively. The quantities $S^{\rm 0}_\lambda$ and 
$S_\lambda$ can be
further expressed as a summation of the corresponding quantities for the disk,
thin disk and bulge:
\begin{eqnarray}
S^{\rm 0}_\lambda & = & S^{\rm 0, disk}_\lambda + S^{\rm 0, tdisk}_\lambda + 
S^{\rm 0, bulge}_\lambda\\
S_\lambda & = & S^{\rm disk}_\lambda + S^{\rm tdisk}_\lambda + 
S^{\rm bulge}_\lambda. 
\end{eqnarray}
The apparent and intrinsic flux densities for the disk, thin disk and bulge are
related as follows:
\begin{eqnarray}
S^{\rm 0, disk}_\lambda & = & S^{\rm disk}_\lambda\,
10^{\frac{\displaystyle {\Delta} m^{\rm disk}_\lambda}{\displaystyle 2.5}}\\
S^{\rm 0, tdisk}_\lambda & = & \frac{S^{\rm tdisk}_\lambda}{1-F\,f_{\lambda}}
10^{\frac{\displaystyle {\Delta} m^{\rm tdisk}_\lambda}{\displaystyle 2.5}}\\
S^{\rm 0, bulge}_\lambda & = & S^{\rm bulge}_\lambda\,
10^{\frac{\displaystyle {\Delta} m^{\rm bulge}_\lambda}{\displaystyle 2.5}},
\end{eqnarray}
where ${\Delta} m^{\rm disk}_\lambda$, ${\Delta} m^{\rm tdisk}_\lambda$ and
${\Delta} m^{\rm bulge}_\lambda$ are the attenuation values for the disk, thin
disk and bulge, which can be derived from Tables 4-6, F is the clumpiness 
factor as defined in Sect.~2.2 and $f_\lambda$ is the
function defined in Appendix A and tabulated in Table~A.1. 
In physical terms,
$F\,f_\lambda$ is the fraction of the emitted UV flux density at wavelength 
$\lambda$ which is locally absorbed in star-forming regions.

If we define the ratios of the apparent luminosities of disk, thin disk and 
bulge to the total apparent luminosity  to be:
\begin{eqnarray}
r^{\rm disk}_\lambda & = & \frac{S^{\rm disk}_\lambda}{S_\lambda}
\\
r^{\rm tdisk}_\lambda & = & \frac{S^{\rm tdisk}_\lambda}{S_\lambda}
\\
r^{\rm bulge}_\lambda & = & \frac{S^{\rm bulge}_\lambda}{S_\lambda},
\end{eqnarray}
then the total attenuation of the galaxy can be expressed as:
\begin{eqnarray}\nonumber
{\Delta} m_\lambda & = & 2.5\log( 
r^{\rm disk}_\lambda\,
10^{\frac{\displaystyle {\Delta} m^{\rm disk}_\lambda}{\displaystyle 2.5}}
\end{eqnarray}
\begin{eqnarray}\nonumber
~~~~~~~~+\, \frac{1-r^{\rm disk}_\lambda-r^{\rm bulge}_\lambda}{1-F\,f_\lambda}\,
10^{\frac{\displaystyle {\Delta} m^{\rm tdisk}_\lambda}{\displaystyle 2.5}}
\end{eqnarray}
\begin{eqnarray}
~~~~~~~~+\, r^{\rm bulge}_\lambda\,
10^{\frac{\displaystyle {\Delta} m^{\rm bulge}_\lambda}{\displaystyle 2.5}}).
\end{eqnarray}
This is a general formula giving the composite attenuation curve as a function
of 5 independent parameters: $i$, $\tau^{\rm f}_{\rm B}$, 
$r^{\rm disk}_{\lambda}$, $r^{\rm bulge}_{\lambda}$ and $F$. For practical
application, this formula can be simplified if we write it separately for the
UV and the optical/NIR ranges. It is a good approximation to consider that 
the thin
disk emits only in the UV range and the disk and bulge only in the optical/NIR
range. This approximation has been validated by our detailed analysis on
NGC~891 in Paper~I. In the UV Eq.~16 becomes:
\begin{eqnarray}
{\Delta} m^{\rm UV}_\lambda & = & {\Delta} m^{\rm tdisk}_{\lambda} - 
2.5\log(1-F\,f_\lambda),
\end{eqnarray}
while in the optical/NIR Eq.~16 becomes:

\begin{eqnarray}\nonumber
{\Delta} m^{\rm optical}_{\lambda} & = & 2.5\log( 
(1-r^{\rm bulge}_\lambda)\,
10^{\frac{\displaystyle {\Delta} m^{\rm disk}_\lambda}{\displaystyle 2.5}} 
\end{eqnarray}
\begin{eqnarray}
~~~~~~~~+\, r^{\rm bulge}_\lambda\,
10^{\frac{\displaystyle {\Delta} m^{\rm bulge}_\lambda}{\displaystyle 2.5}}).
\end{eqnarray}

\subsection{Choice of parameter values}

Both Eqs.~17 and 18 have three free parameters: ${\tau}^{\rm f}_{\rm B}$, 
$i$ and $F$ for the UV range and ${\tau}^{\rm f}_{\rm B}$, $i$ and 
$r^{\rm bulge}_{\lambda}$ for the optical/NIR range. 

For resolved objects, 
the apparent bulge-to-total luminosity ratio $r^{\rm bulge}_{\rm \lambda}$ can 
be derived directly from observations, by decomposing the bulge from the disk.

A realistic derivation of the $F$ factor requires a complete modelling of the
whole UV-submm SED. Another approach would
be to derive the $F$ factor directly from comparing highly resolved FIR images
with images in the UV, which soon may become feasible with the advent of GALEX
in conjunction with ISO and future SIRTF observations of nearby galaxies. For a
simple recipe, the median value of the $F$ factors derived in Paper~II
can be used, namely $F=0.22$, which is also the value obtained for
NGC~891. We note here that it cannot be ruled out that $F$ may increase with
$SFR$, as suggested by the study of the FIR/radio correlation by Pierini et al.
(2003). Physically, this might be expected if an increased SFR is accompanied 
not only by an increase in the number of independent HII regions, but also by 
a higher probability for further star formation to happen preferentially near
already existing HII regions. Then, the $F$ factor would also increase,
as a consequence of the increased blocking capability of the optically
thick molecular clouds in the star-forming complex. 
This would be expected to occur if star formation is a self
propagating phenomenon, in which preceding generations of
stars can trigger the formation of new generations.  

The values of opacities ${\tau}^{\rm f}_{\rm B}$ are also best found from 
a complete modelling of the whole UV-submm SED. In a 
companion paper we will present a grid of calculations for the
FIR/submm emission corresponding to the attenuations presented in Tables 4-6, 
from which ${\tau}^{\rm f}_{\rm B}$ can be extracted from a self-consistent fit
to the entire UV-submm range. In the absence of FIR photometry, an 
alternative method to derive ${\tau}^{\rm f}_{\rm B}$ would be to use the
emission from
the Balmer recombination lines, integrated over the whole galaxy. In our 
model the emission from these lines arises from the thin stellar disk. These
lines will be attenuated both by the diffuse dust (associated with both the
young and old stellar populations) and by the clumpy dust 
(associated with the star-forming regions). Since
in the formulation of our model the 
fraction of the emission locally absorbed in the clumpy
component  is the same for all
lines in the Balmer series, the line ratio will depend only on the 
attenuation by the diffuse dust. This attenuation can readily be found by 
interpolating in wavelength between the attenuation values for the thin disk 
derived from Table~5. As an example, the ratio ${\rm H}\alpha/{\rm H}\beta$ was
derived by interpolating in wavelength between the adjacent optical bands to
the wavelength of the recombination lines. To present the line ratio in an
analogous way to the information from Tables~4-6, we fitted the 
$\log({\rm H}\alpha/{\rm H}\beta)$ with a 5th order polynomial function of the
form:

\begin{eqnarray}
\frac{{\rm H}\alpha}{{\rm H}\beta} = \left\{
  \begin{array}{lll}
   \displaystyle \sum_{{\rm j}=0,n} a_{\rm j}(1-\cos(i))^{\rm j} 
                                    & {\rm ;} & 1-\cos(i) \leq 0.90\\
   b_0                              & {\rm ;} & 1-\cos(i) = 0.95\\
   b_1                              & {\rm ;} & 1-\cos(i) = 1.00\\
  \end{array}
  \right .
\end{eqnarray}
where $n=4$. Values for the coefficients $a_j$ and for the constants $b_0$ 
and $b_1$ are given in Table~7 and the dependence of the 
${\rm H}\alpha/{\rm H}\beta$ line ratio on inclination and 
${\tau}^{\rm f}_{\rm B}$ is shown in Fig.~6. If a measurement of the line 
ratio ${\rm H}\alpha/{\rm H}\beta$ were available for a galaxy of known 
inclination,  this could be compared with the calculated ratios derived from 
Table~7 to derive  ${\tau}^{\rm f}_{\rm B}$. 

\begin{figure}[htb]
\includegraphics[scale=0.7]{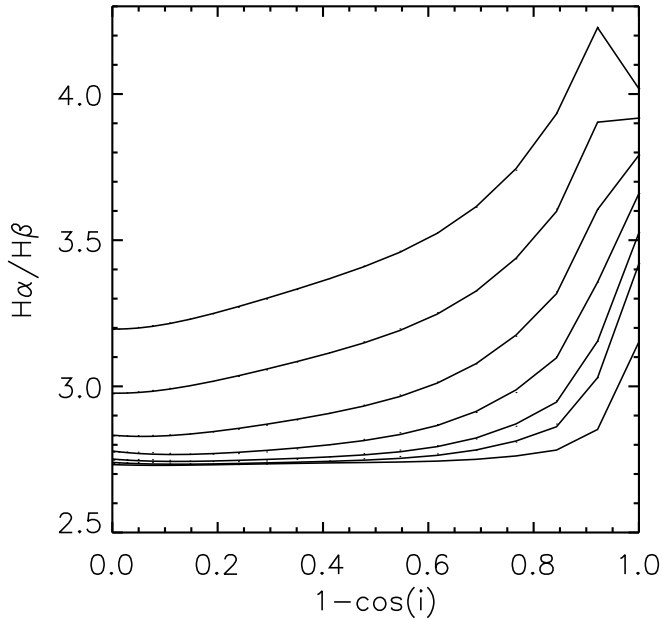}
\caption{The effect of dust on the ${\rm H}\alpha/{\rm H}\beta$ line ratio as 
a function of
inclination $i$. The seven curves (from top to bottom) correspond to 
$\tau^{\rm f}_{\rm B}$: 8, 4, 2, 1, 0.5, 0.3, and 0.1. The 
face-on orientation corresponds to $1-\cos(i) = 0.0$ and the edge-on 
orientation corresponds to  $1-\cos(i) = 1.0$.}
\end{figure} %Fig. 6
\setcounter{table}{6}
\begin{table}[htb]
\include{0689tab7}
\end{table}

\subsection{An example}

To illustrate solutions derived from Eqs.~17-18, we calculate composite 
attenuation curves for NGC~891 as would be obtained for different 
inclinations. For this galaxy ${\tau}^{\rm f}_{\rm B}=4.2$ and $F=0.22$ (see
Paper~I). In this particular case the apparent 
bulge-to-total ratio $r^{\rm bulge}_{\lambda}$ can be derived  from a known 
intrinsic value (0.246, averaged over the optical/NIR; Xilouris et al. 1999) 
by application of Eqns.~10 and 12 for the wavelengths and inclinations of 
interest. Using the values tabulated in Tables~4-6 we obtained the curves 
plotted
in Fig.~7 for three inclinations: 0, 60 and 90 degrees. The curves show a
smooth progression with wavelength, except for the edge-on orientation, where
there is a steep step between optical and UV, due to the large contrast in
attenuation between the thin and thick disks viewed edge-on. 

Using the attenuation curve for NGC~891 ($i=89.8^{\circ}$) and the 
intrinsic flux densities from Paper~I we can predict the apparent
magnitude of the galaxy in the UV and compare this prediction with
observations. Until now the galaxy has been detected only at 2500\,$\AA$
by Marcum et al. (2001) who measured $m^{2500}=13.00\pm0.23$. Our prediction
is $m^{2500}=13.29$, in good agreement with the observations, thus
reinforcing the validity of the model for the intrinsic stellar luminosity 
and attenuation presented in Paper~I. 

\begin{figure}[htb]
\includegraphics[scale=0.7]{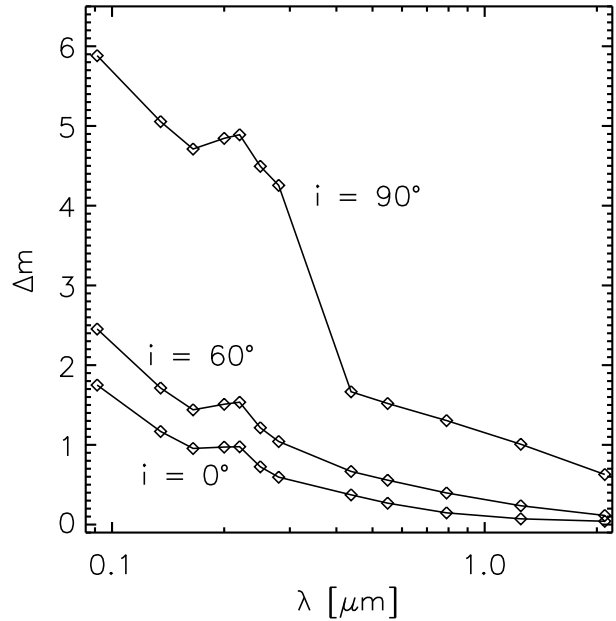}
\caption{Composite attenuation curves for the galaxy NGC~891, as would be
obtained if the galaxy were viewed at inclinations of 0, 60 and 90 degrees. 
$\tau^{\rm f}_{\rm B}=4.2$ and $F=0.22$.}
\end{figure} %Fig. 7

\section{The effect of the bulge-to-disk ratio on the derived opacities}

To illustrate the use of composite attenuation curves in
the interpretation of optical/NIR data we consider the effect of a varying
bulge-to-disk ratio on the inclination dependence of the apparent optical
emission from galaxies. 
Traditionally, astronomers have used this dependence to make statistical
estimates of the opacity of galaxian disks through 
analysis of large samples of spiral galaxies. But these studies have not 
taken into account the fact that the attenuation of the bulge has a different 
variation with inclination than that of the disk, as shown in Sect.~4.

In order to quantify the effect of the bulge-to-disk ratio on the derived
opacities, we calculated the variation of the composite attenuation with
inclination for three different opacities, $\tau^{\rm f}_{\rm B}= 2, 4, 8$. 
These curves were calculated for a grid of values of the bulge-to-disk
ratio ranging from 0 to 100, which represent the whole range between 
``pure disk'' and ``pure bulge'' galaxies. We
chose to do these calculations in the K band in order to compare our results
with recent statistical studies of internal extinction in spiral galaxies
in the NIR. For a convenient comparison with observational studies we fitted
the variation of the composite attenuation with
-$\log(\cos(i))$ using a linear fit\footnote{-$\log(\cos(i))$ is equal to 
$\log(a/b)$, where $a/b$ is the axial ratio of the galaxy.} to the lower 
inclination range ($1-\cos(i)<0.7$). The resulting
slope of the fit $\gamma_{\rm K}$ is plotted in Fig.~8 versus the 
bulge-to-disk
ratio and for the 3 opacities considered. All curves show a ``S-like'' shape,
tending towards the asymptotes of the ``pure'' disk and ``pure'' bulge. 
It can be seen from Fig.~8 that an increase from 0 to 1 in bulge-to-disk 
ratio 
(which embraces the observed range; Trujillo et al. 
2002) produces a comparable 
change in $\gamma_{\rm K}$ to that resulting from a doubling of the opacity of
a ``pure'' disk. For example, changing the bulge-to-disk ratio from 0 to 1 for 
$\tau^{\rm f}_{\rm B}= 2$ increases $\gamma_{\rm K}$ by a factor of 2.2, 
whereas increasing the opacity from $\tau^{\rm f}_{\rm B}= 2$ to 4 changes 
$\gamma_{\rm K}$ by a factor of only 1.9 for a ``pure'' disk galaxy. {\it 
Thus, 
increasing the bulge-to-disk ratio at a constant opacity can mimic the effect 
of increasing the opacity of a ``pure'' disk.} 

\begin{figure}[htb]
\includegraphics[scale=0.68]{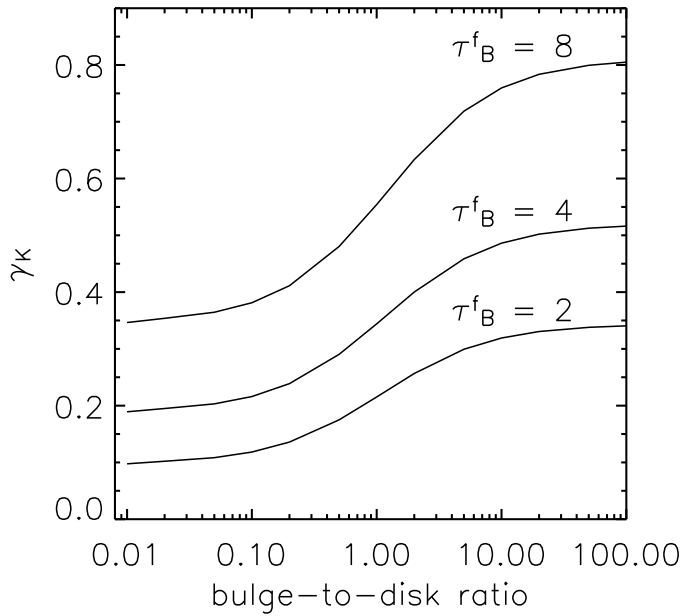}
\caption{The slope $\gamma_{\rm K}$ versus the intrinsic 
bulge-to-disk ratio, where  $\gamma_{\rm K}$ is the slope
of a linear fit to the low inclination part ($1-\cos(i) < 0.7$) of the 
composite attenuation curve (${\Delta}$m vs -$\log(\cos(i)$)) in the K 
band. The calculations were performed for $\tau^{\rm f}_{\rm B}=8, 4, 2$.}
\end{figure} %Fig. 8

This may have
consequences for the use of optical statistical samples to evaluate the
dependence of opacity on galaxy luminosity. For example, in their study of 
15224 spiral 
galaxies from the 2 Micron All-Sky Survey, Masters et al. (2003) found a trend 
for the slope $\gamma_{\rm K}$ (and also for the corresponding slopes in J and
H bands) to increase with increasing K-band luminosity. These authors interpret
this result as a trend of increasing disk opacity with increasing K-band 
luminosity. However, galaxies that
 have brighter K-band luminosities are 
also biased towards earlier-type spirals (e.g. Boselli et al. 1997), which in 
turn have larger bulge-to-disk ratios. Fig.~13 of Masters et al. (2003) show a
variation of 0.2 in $\gamma_{\rm K}$ over the luminosity range of their
sample, which is comparable to the variation in $\gamma_{\rm K}$ found in our
Fig.~8 between bulge-to-disk ratios of 0 and 1. This raises
the possibility that some of the trend found by Masters et al. (2003) is due 
to a systematic increase in the bulge-to-disk ratio with luminosity in their 
sample. It will be important to quantify this effect in studies that 
investigate the dependence of the 
ratio of the FIR luminosity to the intrinsic UV luminosity of gas rich
galaxies on the stellar mass of the galaxies (Pierini \& M\"oller 2003).
{\it In
general, ignoring the presence of bulges can lead to a systematic
overestimate of the opacity of disks.}

\section{Summary and conclusions}

We present new calculations for the attenuation of the integrated 
stellar light from spiral galaxies, utilising geometries for stars and dust 
constrained by a joint consideration of the UV/optical and FIR/submm SEDs. In 
addition to the single exponential diffuse dust disk used in previous studies 
of attenuation, we also invoke a clumpy and strongly 
heated distribution of grains spatially correlated with the star-forming
regions (required to account for the FIR colours) and diffuse dust 
associated with the spiral arms (needed to account for the amplitude of the 
submm emission). The latter is approximated by a second exponential
dust disk having the same spatial distribution as the young, UV emitting 
stellar population, also approximated by an exponential disk - the 
``thin disk''. The old, optical/NIR-emitting stellar population is specified 
by an exponential disk of larger scale height - the ``disk'',  plus a de 
Vaucouleurs bulge - the ``bulge''.

Radiation transfer calculations were performed separately for the three 
diffuse emissivity components (disk, thin disk and bulge) seen through the 
same fixed distribution of diffuse dust. We used the radiative transfer code of
Kylafis \& Bahcall (1987) which includes anisotropic multiple scattering. The
attenuation of each component was calculated for a grid of central 
face-on B-band optical depth  $\tau^{\rm f, disk}_{\rm B}$ and inclination
$i$, for a range of wavelengths from 912\,$\AA$ to 2.2\,${\mu}$m.
The resulting curves of attenuation versus inclination were fitted with
polynomial function and the coefficients tabulated in Tables 4-6. The tables
allow the attenuation to be obtained for any desired combination of $\tau^{\rm
f, disk}_{\rm B}$, $i$, and wavelength, by interpolation in 
$\tau^{\rm f, disk}_{\rm B}$ and wavelength.

The clumpy component was handled analytically, whereby the wavelength 
dependence of local absorption of starlight in star-forming regions was purely
determined from geometrical considerations.
By combining the global 
attenuation of the individual stellar populations from the disk, thin disk 
and bulge due to the diffuse dust with the local attenuation due to the 
clumpy dust associated with the star-forming regions, we obtain a general 
formula for the calculation of composite attenuation of
the integrated emission from spiral galaxies. 
As an example of this formula, we calculated composite attenuation curves for 
NGC~891 as would be seen at different inclinations. Using the
attenuation curve for NGC~891 at the actual inclination $i=89.8^{\circ}$ we 
predicted the apparent magnitude of this galaxy in the UV and found this
prediction to be in good agreement with the recent observations of Marcum et 
al. (2001). 

We also used our model to derive the ratio ${\rm H}\alpha/{\rm H}\beta$ as a
function of inclination and $\tau^{\rm f}_{\rm B}$. This information is
presented in Table~7 in form of polynomial fits and can be used to derive the
value of $\tau^{\rm f}_{\rm B}$ for objects with no FIR/submm photometry.

The detailed analysis of the attenuation properties of the individual
geometrical components led to the following conclusions:
\begin{itemize}
\item For a typical galaxy 
and in the optical/NIR spectral range, the relative 
attenuation between the disk, thin disk and bulge depends strongly on 
inclination. At low and 
intermediate inclinations, the bulge suffers a larger attenuation than
  that of the disk or thin disk, and has the steepest dependence on 
wavelength. Furthermore, in this regime, 
the curves of attenuation versus wavelength for the disk and thin disk show 
quite similar behaviour to each other, indicating that the attenuation is not 
sensitive to the differences in scale height and scale length between the two 
disks.
At higher inclinations, however, the attenuation curves 
for the thin disk start to diverge from those of the disk. For the edge-on 
orientation the attenuation of the thin disk dominates that of the other 
geometrical components, reversing the situation for the face-on orientation, 
where the thin disk has the smallest attenuation.
\item
Increasing the bulge-to-disk ratio at a constant opacity can mimic
the effect of increasing the opacity of a ``pure'' disk. In
general, ignoring the presence of bulges can lead to a systematic
overestimate of the opacity of disks. 
\end{itemize}

\acknowledgements
M. Dopita acknowledges the support of the Australian National University
and of the Australian Research Council through his ARC Australian Federation
Fellowship and through his ARC Discovery project DP0208445.
We would like to thank our referee, S. Bianchi, for his insightful and helpful
 comments and suggestions, which helped improve the paper.

\setcounter{table}{3}
\begin{table}[htb]
\include{0689t456}
\end{table}
\appendix

\section{Wavelength dependence of local absorption of starlight in star-forming
regions}

As introduced by Popescu et al. (2000), the clumpiness factor $F$ denotes the 
total fraction of UV light which is locally absorbed in the star-forming 
regions where the stars were born. In the current work we need to understand
the wavelength dependence of the probability for local absorption of UV
photons, which we denote by $F_{\lambda}$, such that
\begin{eqnarray}
F = \frac{
     \int_{\lambda 1}^{\lambda 2} F_\lambda L_{\lambda} d{\lambda}
     }
     {
     \int_{\lambda 1}^{\lambda 2} L_{\lambda} d{\lambda}
      }
\end{eqnarray}
where $L_{\lambda}$ is the intrinsic luminosity density of the galaxy and
$\lambda 1$ and $\lambda 2$ define the UV spectral range.
In our formulation $F_{\lambda}$ is 
determined only by
geometric considerations. For a given star, the probability of local absorption
of a photon is simply proportional to the solid angle of the parent
cloud subtended at the star. The cloud is considered to be so opaque that a
photon is absorbed independent of its wavelength if it passes through the
cloud material. The wavelength dependence arises because stars of different 
masses survive for different times, such that lower mass and redder stars can 
escape further from the star-forming regions in their lifetimes.

We approximate the solid angle $\Omega$ of a parent cloud subtended at an
offspring star of age t with:
\begin{eqnarray}
\Omega = \left\{
\begin{array}{lll}
4\pi\,p_0 & {\rm for} & t<t_{\rm local},\\
2\pi\,p_0[1-(1-(\frac{\displaystyle t_{\rm local}}{\displaystyle t})^2)^{1/2}]
      & {\rm for} & t \ge t_{\rm local}\\
\end{array}
\right.
\end{eqnarray}
where $p_0$ is the ``porosity factor'' giving the typical fraction of those 
lines of sight passing through the cloud which are blocked by the cloud 
material and $t_{\rm local}$ is 
the typical time taken for the star to escape from the star-forming region.
This simply approximates the star-forming regions with porous spheres. The
porosity of the spheres, averaged over the whole galaxy, is taken to be the 
same for all stellar masses, so that $p_0$ is independent of M. A further
approximation is that each star is
taken to be born at the centre of the cloud and move outwards at a constant
velocity. 

The expectation value for the fraction of stellar light blocked by
star-forming regions for stars of zero-age main sequence (ZAMS) mass M,
averaged over the lifetime $t_{\rm gal}$ of the galaxy is:
\begin{eqnarray}
p(M) & = & \frac{
            \displaystyle \int^{t_{\rm gal}}_{t_0(M)} 
           \Psi(t_{\rm gal}-t)\,
           \Phi(M)\,dt\,\frac{\displaystyle \Omega(t_{\rm gal}-t)}{4\pi}
            }
            {
            \displaystyle \int^{t_{\rm gal}}_{t_0(M)} 
            \Psi(t_{\rm gal}-t)\,\Phi(M)\,dt
            }
\end{eqnarray}
where $\Phi$ is the initial mass function (IMF), $\Psi$ is the star-formation 
history of the galaxy and 
\begin{eqnarray}
t_0(M)=\left\{
    \begin{array}{lll}
     t_{\rm gal}-t_{\rm star}(M) & {\rm for} & t_{\rm star}(M) < t_{\rm gal}\\
     0                    & {\rm for} & t_{\rm star}(M) \ge t_{\rm gal}\\
    \end{array}
    \right .
\end{eqnarray}
where $t_{\rm star}(M)$ is the lifetime of a star of ZAMS mass M.
Here we consider a Salpeter IMF and an exponentially declining star-formation
rate:
\begin{eqnarray}
\Psi(t) = SFR \times e^{t/t_{\rm c}}
\end{eqnarray}
where $SFR$ is the current star-formation rate and $t_{\rm c}$ is the time 
constant.

The power density $P_{\lambda}(M)$ from a star with ZAMS mass M is given by:
\begin{eqnarray}
P_\lambda(M) = \int^{t_{\rm gal}}_{t_0(M)}SED_\lambda(M,t)\,
                \Psi(t_{\rm gal}-t)\,\Phi(M)\,dt
\end{eqnarray}
Here we approximate the spectral energy distribution 
$SED_\lambda(M,t)$ from an individual star with a time independent black-body 
function:
\begin{eqnarray}
SED_\lambda(M) = \frac{
               B_\lambda(T(M))
               }
               {
               \int B_\lambda(T(M))\,d{\lambda}
               }
               L(M)
\end{eqnarray}
where $L(M)$ and T(M) represent the ZAMS luminosity and temperature of a star 
of mass M.

We derived $L(M)$ from Maeder (1987) and Smith (1983):
\begin{eqnarray}
\frac{L}{L_{\odot}} = \left\{
 \begin{array}{lll}
  26.3\,(\frac{\displaystyle M}{\displaystyle M_{\odot}})^{1.84} & 
  {\rm for} & \frac{\displaystyle M}{\displaystyle M_{\odot}} \ge 40\\\\
  18.6\,(\frac{\displaystyle M}{\displaystyle M_{\odot}})^{2.55} & 
  {\rm for} & 7.5 \le \frac{\displaystyle M}{\displaystyle M_{\odot}} < 40\\\\
  1.0\,(\frac{\displaystyle M}{\displaystyle M_{\odot}})^{4.0} & 
  {\rm for} & 0.4 \le \frac{\displaystyle M}{\displaystyle M_{\odot}} < 7.5\\\\
  0.23\,(\frac{\displaystyle M}{\displaystyle M_{\odot}})^{2.3} & 
  {\rm for} & \frac{\displaystyle M}{\displaystyle M_{\odot}} < 0.4\\
 \end{array}
 \right .
\end{eqnarray}
T(M) was derived from data in Allen (2000), Lang(1980) and Schaerer et al. 
(1996):
\begin{eqnarray}
\log T(M) = \sum_{i=0,5} a_i\,[\log (\frac{M}{M_{\odot}})]^i
\end{eqnarray}
where $a_i=[3.756,0.601,0.250,-0.147,-0.078,0.038]$.
The dependence of stellar main sequence lifetime on ZAMS mass M was derived
from Maeder (1987), Sandage (1957) and Smith (1983):
\begin{eqnarray}
\frac{t_{\rm star}}{10^7 \rm yr} = \left\{
 \begin{array}{lll}
 2.19\,(\frac{\displaystyle M}{\displaystyle M_{\odot}})^{-0.430} & 
 {\rm for} & \frac{\displaystyle M}{\displaystyle M_{\odot}} \ge 40\\\\
 15.4\,(\frac{\displaystyle M}{\displaystyle M_{\odot}})^{-0.963} & 
 {\rm for} & 8.1 \le \frac{\displaystyle M}{\displaystyle M_{\odot}} < 40\\\\
 1100\,(\frac{\displaystyle M}{\displaystyle M_{\odot}})^{-3.0} & 
 {\rm for} & 0.4 \le \frac{\displaystyle M}{\displaystyle M_{\odot}} < 8.1\\\\
 4800\,(\frac{\displaystyle M}{\displaystyle M_{\odot}})^{-1.3} & 
  {\rm for} & \frac{\displaystyle M}{\displaystyle M_{\odot}} < 0.4\\
 \end{array}
 \right .
\end{eqnarray}

\begin{figure}[htb]
\includegraphics[scale=0.7]{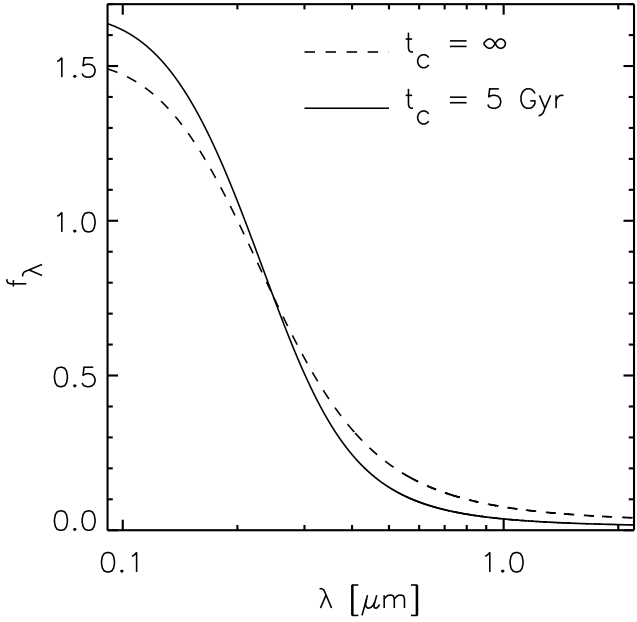}
\caption{The wavelength dependence of the function $f_{\lambda}$. The
solid line indicates the solution for an exponential declining star-formation
rate with $t_{\rm c}=5\,$Gyr. The dashed 
line indicates the solution for a constant star-formation rate 
$t_{\rm c}=\infty$. Both curves were 
calculated for $t_{\rm gal}=10$\,Gyr and $t_{\rm local}=3\times 10^7\,$yr.}
\end{figure}

$F_\lambda$ is then given by the luminosity weighted 
average of $p(M)$ over the stellar mass distribution:
\begin{eqnarray}
F_{\lambda} = \frac{
              \int P_{\lambda}(M)\,p(M)\,dM
              }
              {
              \int P_{\lambda}(M)\,dM
              }
\end{eqnarray}
For normal galaxies $F_{\lambda}$ is completely specified by the porosity 
factor $p_0$, $t_{\rm local}$, $t_{\rm c}$ and $t_{\rm gal}$. Since
\begin{eqnarray}
F_{\lambda}=p_0\,F_{\lambda}(p_0=1)
\end{eqnarray}
Eq. A.1 can be rewritten in terms of $p_0$ as  
\begin{eqnarray}
F = \frac{
     p_0\,\int_{\lambda 1}^{\lambda 2} F_\lambda(p_0=1) L_{\lambda} d{\lambda}
     }
     {
     \int_{\lambda 1}^{\lambda 2} L_{\lambda} d{\lambda}
      }
\end{eqnarray}
Thus, if we define a function $f_{\lambda}$ by 
\begin{eqnarray}
f_{\lambda} = F_{\lambda}(p_0=1)\times
      \frac{
      \int_{\lambda 1}^{\lambda 2} L_{\lambda} d{\lambda}
      }
      {
      \int_{\lambda 1}^{\lambda 2} F_\lambda(p_0=1) L_{\lambda} d{\lambda}
      }
\end{eqnarray}
the clumpiness factor $F$ can be expressed as:
\begin{eqnarray}
F = \frac{F_{\lambda}}{f_{\lambda}}
\end{eqnarray}

It now remains to calculate $f_{\lambda}$, which is a function of only three 
parameters: $t_{\rm local}$, $t_{\rm c}$ and $t_{\rm gal}$. 
For a typical velocity of a star
of 5\,km/s and dimension of star-formation complex of 150\,pc, $t_{\rm local}$
is about $3\times 10^7$\,yr, which is adopted here. For $t_{\rm gal}$ and 
$t_{\rm c}$ we take 10 and 5\,Gyr, respectively. The resulting solution 
$f_{\lambda}$ is tabulated in Table.~A.1, as a function of wavelength and is
plotted as a solid line in Fig.~A.1. The porosity factor corresponding to
$F=0.22$ is
$p_0=0.368$. The assumptions about the star-formation
history contained in $t_{\rm c}$ do not play an important role in shaping the
$f_{\lambda}$. To check this we also calculated $f_{\lambda}$ for the same
parameters but $t_{\rm c}=\infty\,$. This is plotted as the dotted line in
Fig.~A.1. Indeed, there seems to be little difference between the two curves.
This is to be expected, bearing in mind that the timescale for a star to
move out of the star-forming region is much less than the age of the galaxy.
In this paper we adopt the solution for $t_{\rm c}=5\,$Gyr, which
is the value used for NGC~891 in Popescu et al. (2000). 

\begin{table}[htb]
\caption{The wavelength dependence of the function $f_{\lambda}$ defined by
Eq. A.~13.}
\begin{tabular}{lc}
\hline
$\lambda$ & $f_{\lambda}$\\
$\AA$     & -            \\
\hline
  912 & 1.636\\
 1350 & 1.473\\
 1650 & 1.300\\
 2000 & 1.064\\
 2200 & 0.927\\
 2500 & 0.745\\
 2800 & 0.591\\
 4400 & 0.191\\
 5500 & 0.109\\
 9000 & 0.045\\
12500 & 0.027\\
22000 & 0.018\\
\hline
\end{tabular}
\end{table}
\end{document}

%% file: 0689tab7.tex
\caption{Coefficients and constants for the calculation of the 
${\rm H}\alpha/{\rm H}\beta$ line ratio}
\begin{tabbing}
ss \= sssss \= sssssss \= ssssssss \= sssssss \= ssssssss \= sssssss 
\= sssssss \= sssssss \= sssssss \kill
\> \hspace{0.05cm}$\tau^{\rm f}_{\rm B}$
     \> \hspace{0.3cm} $a_0$  \>  \hspace{0.45cm} $a_1$ \>  
\hspace{0.4cm}$a_2$ \>  \hspace{0.55cm}$a_3$ \> \hspace{0.4cm}$a_4$ \> 
\hspace{0.45cm}$b_0$ \>  \hspace{0.45cm}$b_1$\\ 
\> 0.1\>  2.732\> -0.062\>  0.472\> -0.989\>  0.711\>  2.853\>  3.152\\
\> 0.3\>  2.740\> -0.118\>  0.716\> -1.477\>  1.181\>  3.030\>  3.419\\
\> 0.5\>  2.751\> -0.159\>  0.963\> -1.967\>  1.629\>  3.155\>  3.525\\
\> 1.0\>  2.779\> -0.237\>  1.421\> -2.638\>  2.155\>  3.357\>  3.660\\
\> 2.0\>  2.833\> -0.167\>  1.721\> -3.207\>  2.619\>  3.605\>  3.791\\
\> 4.0\>  2.977\> -0.044\>  1.945\> -3.728\>  2.988\>  3.904\>  3.918\\
\> 8.0\>  3.196\> -0.042\>  2.435\> -4.730\>  3.709\>  4.227\>  4.018\\
\end{tabbing}

%% file: 0689t456.tex
\caption{Coefficients and constants for the calculation of the attenuation of
the disk}
\begin{tabbing}
\= sssss \= ssssssss \= ssssssss \= sssssss \= sssssss \= ssssssss 
\= sssssss \= sssssss \= sssssss \kill
\>   \hspace{0.05cm}$\tau^{\rm f}_{\rm B}$
     \>\hspace{0.4cm} $a_0$  \>  \hspace{0.55cm}$a_1$ \> \hspace{0.4cm} $a_2$
     \> \hspace{0.4cm} $a_3$ \>  \hspace{0.65cm}$a_4$ \> \hspace{0.5cm}$a_5$ 
     \> \hspace{0.45cm} $b_0$ \>  \hspace{0.5cm}$b_1$\\ 
\>\hspace{0.35cm}B band\\
\>0.1 \> -0.002 \> \,\,0.069 \> -0.725 \> \,\,2.645 \> \,\,\,-3.883 \> \,\,2.107
\> \,\,0.163 \> \,\,0.240\\
\>0.3 \> -0.003 \> \,\,0.038 \> -0.458 \> \,\,2.528 \> \,\,\,-4.536 \> \,\,2.935
\> \,\,0.390 \>  \,\,0.483\\
\>0.5 \> \,\,0.008 \> \,\,0.055 \> -0.785 \> \,\,3.964 \> \,\,\,-6.569 \> \,\,4.037
\>\,\,0.544 \> \,\,0.632\\
\>1.0 \> \,\,0.022 \> \,\,0.210 \> -1.352 \> \,\,5.519 \> \,\,\,-8.182 \> \,\,4.799
\>\,\,0.795 \> \,\,0.874\\
\>2.0 \> \,\,0.083 \> \,\,0.321 \> -1.493 \> \,\,6.148 \> \,\,\,-8.757 \> \,\,5.039
\>\,\,1.086 \> \,\,1.165\\
\>4.0 \> \,\,0.215 \> \,\,0.491 \> -1.878 \> \,\,7.473 \>-10.356 \> \,\,5.757
\> \,\,1.414 \>  \,\,1.505\\
\>8.0 \> \,\,0.432 \> \,\,0.635 \> -2.166 \> \,\,8.304 \>-11.297 \> \,\,6.189
\>\,\,1.783 \> \,\,1.892\\
\>\hspace{0.35cm}V band\\
\>0.1 \> -0.005 \> \,\,0.060 \> -0.636 \> \,\,2.303 \> \,\,\,-3.346 \> \,\,1.786
\> \,\,0.124 \> \,\,0.191\\ 
\>0.3 \> -0.006 \> \,\,0.033 \> -0.462 \> \,\,2.371 \> \,\,\,-4.125 \> \,\,2.596
\> \,\,0.317 \> \,\,0.408\\
\>0.5 \> -0.002 \> \,\,0.046 \> -0.628 \> \,\,3.265 \> \,\,\,-5.552 \> \,\,3.463
\>\,\,0.456 \> \,\,0.544\\
\>1.0 \> \,\,0.007 \> \,\,0.125 \> -0.884 \> \,\,4.200 \> \,\,\,-6.720 \>\,\,4.158
\>\,\,0.689 \> \,\,0.769\\
\>2.0 \> \,\,0.048 \> \,\,0.229 \> -1.048 \> \,\,4.815 \> \,\,\,-7.218 \> \,\,4.378
\> \,\,0.966 \> \,\,1.040\\
\>4.0 \> \,\,0.148 \> \,\,0.406 \> -1.518 \> \,\,6.391 \> \,\,\,-9.025 \> \,\,5.149
\>\,\,1.281 \> \,\,1.361\\
\>8.0 \> \,\,0.333 \> \,\,0.582 \> -1.996 \> \,\,7.855 \>-10.787 \>\,\,5.947
\>\,\,1.632 \> \,\,1.730\\
\>\hspace{0.35cm}I band\\ 
\>0.1 \> -0.006 \> \,\,0.033 \> -0.385 \> \,\,1.515 \> \,\,\,-2.289 \> \,\,1.239
\>\,\,0.080 \> \,\,0.130\\
\>0.3 \> -0.005 \> \,\,0.063 \> -0.752 \> \,\,2.996 \> \,\,\,-4.546 \> \,\,2.535
\>\,\,0.223 \> \,\,0.306\\
\>0.5 \> -0.005 \> \,\,0.056 \> -0.718 \> \,\,3.215 \> \,\,\,-5.139 \> \,\,3.025
\>\,\,0.336 \> \,\,0.424\\
\>1.0 \> -0.002 \> -0.029 \> \,\,0.164 \> \,\,1.111 \> \,\,\,-3.019 \> \,\,2.460
\>\,\,0.541 \> \,\,0.621\\
\>2.0 \> \,\,0.018 \> \,\,0.046 \> \,\,0.087 \> \,\,1.456 \> \,\,\,-3.303 \> 
\,\,2.690 \>\,\,0.797 \> \,\,0.864\\
\>4.0 \> \,\,0.077 \>	\,\,0.201 \> -0.224 \>  \,\,2.368 \> \,\,\,-4.028 \> \,\,2.925 \> \,\,1.092 \> \,\,1.156\\
\>8.0 \> \,\,0.210 \> \,\,0.427 \> -0.931 \> \,\,4.499 \> \,\,\,-6.498 \> \,\,3.975
\>\,\,1.422 \> \,\,1.497\\
\>\hspace{0.35cm}J band\\ 
\>0.1 \> -0.004 \> \,\,0.020 \> -0.237 \> \,\,0.924 \> \,\,\,-1.353 \> \,\,0.699
\>\,\,0.038 \> \,\,0.065\\
\>0.3 \> -0.004 \> \,\,0.015 \> -0.245 \> \,\,1.300 \> \,\,\,-2.241 \> \,\,1.329
\> \,\,0.119 \> \,\,0.177\\
\>0.5 \> -0.004 \> \,\,0.001 \> -0.121 \> \,\,1.144 \> \,\,\,-2.308 \>  \,\,1.529
\> \,\,0.190 \> \,\,0.264\\
\>1.0 \> -0.002 \> \,\,0.023 \> -0.234 \> \,\,1.835 \> \,\,\,-3.541 \> \,\,2.347
\> \,\,0.339 \> \,\,0.420\\
\>2.0 \> \,\,0.008 \> \,\,0.074 \> -0.409 \> \,\,2.667 \> \,\,\,-4.827 \> \,\,3.189
\> \,\,0.551 \> \,\,0.620\\
\>4.0 \> \,\,0.038 \> \,\,0.134 \> -0.335 \> \,\,2.571 \> \,\,\,-4.564 \> \,\,3.173
\>\,\,0.814 \> \,\,0.865\\
\>8.0 \> \,\,0.112 \> \,\,0.252 \> -0.362 \> \,\,2.655 \> \,\,\,-4.250 \> \,\,2.934
\> \,\,1.115 \> \,\,1.159\\
\>\hspace{0.35cm}K band\\ 
\>0.1 \> -0.003 \> \,\,0.024 \> -0.256 \> \,\,0.832 \> \,\,\,-1.037 \> \,\,0.452
\> \,\,0.009 \> \,\,0.018\\
\>0.3 \> -0.001 \> \,\,0.033 \> -0.357 \> \,\,1.266 \> \,\,\,-1.744 \> \,\,0.856
\>\,\,0.039 \> \,\,0.065\\
\>0.5 \> \,\,0.000 \> \,\,0.038 \> -0.433 \> \,\,1.622 \> \,\,\,-2.336 \> \,\,1.200
\> \,\,0.068 \> \,\,0.107\\
\>1.0 \> \,\,0.006 \> -0.045 \> \,\,0.185 \> \,\,0.138 \> \,\,\,-0.903 \> \,\,0.787
\> \,\,0.137 \> \,\,0.196\\
\>2.0 \> \,\,0.013 \> -0.003 \> -0.096 \> \,\,1.225 \> \,\,\,-2.544 \> \,\,1.724
\> \,\,0.256 \> \,\,0.329\\
\>4.0 \> \,\,0.029 \> \,\,0.055 \> -0.407 \> \,\,2.446 \> \,\,\,-4.369 \> \,\,2.801
\>\,\,0.441 \> \,\,0.508\\
\>8.0 \> \,\,0.063 \> \,\,0.124 \> -0.545 \> \,\,3.010 \> \,\,\,-5.097 \> \,\,3.309
\>\,\,0.688 \> \,\,0.731\\
\end{tabbing}
\end{table}
\begin{table}[htb]
\caption{Coefficients and constants for the calculation of the attenuation of
the thin disk}
\begin{tabbing}
\= ssss \= sssssss \= ssssssss \= sssssss \= sssssssss \= ssssssss 
\= ssssssss \= sssssss \= sssssss \kill
\>   \hspace{0.05cm}$\tau^{\rm f}_{\rm B}$
     \>\hspace{0.4cm} $a_0$  \>  \hspace{0.55cm}$a_1$ \> \hspace{0.4cm} $a_2$
     \> \hspace{0.65cm}$a_3$ \>  \hspace{0.7cm}$a_4$ \> \hspace{0.65cm}$a_5$ 
     \> \hspace{0.45cm} $b_0$ \>  \hspace{0.5cm}$b_1$\\ 
\>\hspace{0.35cm}UV 912\\
\> 0.1 \> \,\,0.046 \> \,\,0.226 \> -2.242\> \,\,\,\,\,9.055\> -14.193\>
\,\,\,\,\,8.142 \> \,\,0.884\> \,\,1.677\\ 
\> 0.3 \> \,\,0.164 \> \,\,0.219 \> -3.191\> \,\,15.257\> -24.502\> \,\,14.129
\> \,\,1.719\> \,\,2.622\\ 
\> 0.5 \> \,\,0.235 \> \,\,0.353 \> -2.841\> \,\,14.431\> -23.840\> \,\,14.305
\> \,\,2.228\> \,\,3.120\\ 
\> 1.0 \> \,\,0.440 \> \,\,0.445 \> -2.370\> \,\,14.574\> -25.276\> \,\,15.771
\> \,\,3.044\> \,\,3.843\\ 
\> 2.0 \> \,\,0.773 \> \,\,0.509 \> -1.292\> \,\,12.835\> -24.355\> \,\,16.239
\> \,\,3.979\> \,\,4.604\\ 
\> 4.0 \> \,\,1.250 \> \,\,0.450 \> \,\,1.412\> \,\,\,\,\,5.345\> -15.012\> 
\,\,12.461\> \,\,4.973\> \,\,5.379\\ 
\> 8.0 \> \,\,1.891 \> \,\,0.165 \> \,\,6.978\> -12.176\> \,\,\,\,\,9.259\> 
\,\,\,\,\,0.833\> \,\,5.956\> \,\,6.136\\ 								     
\>\hspace{0.35cm}UV 1350\\						     
\> 0.1 \> -0.001 \> \,\,0.165 \> -1.593\> \,\,\,\,\,6.421\> -10.241\> 
\,\,\,\,\,5.876\> \,\,0.532\> \,\,1.198\\ 
\> 0.3 \> \,\,0.057 \> \,\,0.246 \> -3.086\> \,\,13.069\> -20.582\> \,\,11.719
\> \,\,1.158\> \,\,2.022\\ 
\> 0.5 \> \,\,0.095 \> \,\,0.272 \> -3.018\> \,\,13.642\> -21.967\> \,\,12.875
\> \,\,1.566\> \,\,2.467\\ 
\> 1.0 \> \,\,0.197 \> \,\,0.351 \> -2.862\> \,\,14.467\> -23.967\> \,\,14.492
\> \,\,2.244\> \,\,3.131\\ 
\> 2.0 \> \,\,0.396 \> \,\,0.453 \> -2.434\> \,\,14.793\> -25.590\> \,\,16.012
\> \,\,3.065\> \,\,3.852\\ 
\> 4.0 \> \,\,0.730 \> \,\,0.627 \> -1.623\> \,\,13.302\> -24.661\> \,\,16.376
\> \,\,4.000\> \,\,4.610\\ 
\> 8.0 \> \,\,1.226 \> \,\,0.574 \> \,\,1.049\> \,\,\,\,\,5.861\> -15.294\> 
\,\,12.527\> \,\,4.989\> \,\,5.377\\ 								     
\>\hspace{0.35cm}UV 1650\\						 
\> 0.1 \> -0.011 \> \,\,0.149 \> -1.426\> \,\,\,\,\,5.731\> \,\,\,-9.137\> 
\,\,\,\,\,5.209\> \,\,0.434\> \,\,1.044\\ 
\> 0.3 \> \,\,0.030 \> \,\,0.264 \> -3.064\> \,\,12.364\> -19.204\> \,\,10.828
\> \,\,0.984\> \,\,1.820\\ 
\> 0.5 \> \,\,0.061 \> \,\,0.249 \> -3.109\> \,\,13.590\> -21.632\> \,\,12.502
\> \,\,1.356\> \,\,2.244\\ 
\> 1.0 \> \,\,0.131 \> \,\,0.313 \> -2.979\> \,\,14.358\> -23.483\> \,\,14.044
\> \,\,1.981\> \,\,2.882\\ 
\> 2.0 \> \,\,0.289 \> \,\,0.430 \> -2.806\> \,\,15.516\> -26.151\> \,\,15.996
\> \,\,2.752\> \,\,3.582\\ 
\> 4.0 \> \,\,0.579 \> \,\,0.579 \> -2.043\> \,\,14.310\> -25.701\> \,\,16.608
\> \,\,3.649\> \,\,4.327\\ 
\> 8.0 \> \,\,1.022 \> \,\,0.545 \> \,\,0.171\> \,\,\,\,\,8.751\> -19.373\>
\,\,14.399\> \,\,4.624\> \,\,5.089\\ 								     
\>\hspace{0.35cm}UV 2000\\						     
\> 0.1 \> -0.012 \> \,\,0.114 \> -1.051\> \,\,\,\,\,4.721\> \,\,\,-8.055\> 
\,\,\,\,\,4.873\> \,\,0.510\> \,\,1.163\\ 
\> 0.3 \> \,\,0.032 \> \,\,0.249 \> -2.938\> \,\,12.319\> -19.339\> \,\,11.043
\> \,\,1.116\> \,\,1.974\\ 
\> 0.5 \> \,\,0.067 \> \,\,0.246 \> -3.087\> \,\,14.072\> -22.503\> \,\,13.052
\> \,\,1.514\> \,\,2.413\\ 
\> 1.0 \> \,\,0.151 \> \,\,0.321 \> -2.833\> \,\,14.546\> -24.011\> \,\,14.433
\> \,\,2.179\> \,\,3.068\\ 
\> 2.0 \> \,\,0.337 \> \,\,0.462 \> -2.345\> \,\,14.461\> -25.125\> \,\,15.748
\> \,\,2.985\> \,\,3.783\\ 
\> 4.0 \> \,\,0.671 \> \,\,0.587 \> -1.559\> \,\,13.289\> -24.689\> \,\,16.342
\> \,\,3.909\> \,\,4.537\\ 
\> 8.0 \> \,\,1.158 \> \,\,0.498 \> \,\,1.246\> \,\,\,\,\,5.691\> -15.410\>
\,\,12.644\> \,\,4.892\> \,\,5.300\\ 								     
\>\hspace{0.35cm}UV 2200\\						     
\> 0.1 \> -0.001 \> \,\,0.128 \> -1.131\> \,\,\,\,\,4.768\> \,\,\,-7.887\>
\,\,\,\,\,4.745\> \,\,0.550\> \,\,1.217\\ 
\> 0.3 \> \,\,0.042 \> \,\,0.201 \> -2.522\> \,\,11.297\> -18.234\> \,\,10.643
\> \,\,1.177\> \,\,2.041\\ 
\> 0.5 \> \,\,0.077 \> \,\,0.222 \> -2.732\> \,\,13.130\> -21.415\> \,\,12.632
\> \,\,1.584\> \,\,2.485\\ 
\> 1.0 \> \,\,0.168 \> \,\,0.310 \> -2.513\> \,\,13.748\> -23.141\> \,\,14.120
\> \,\,2.263\> \,\,3.147\\ 
\> 2.0 \> \,\,0.367 \> \,\,0.467 \> -2.108\> \,\,13.864\> -24.484\> \,\,15.536
\> \,\,3.082\> \,\,3.867\\ 
\> 4.0 \> \,\,0.716 \> \,\,0.590 \> -1.304\> \,\,12.625\> -23.923\> \,\,16.059
\> \,\,4.014\> \,\,4.623\\ 
\> 8.0 \> \,\,1.219 \> \,\,0.476 \> \,\,1.751\> \,\,\,\,\,4.159\> -13.387\>
\,\,11.725\> \,\,4.999\> \,\,5.387\\ 					     
\>\hspace{0.35cm}UV 2500\\						     
\> 0.1 \> -0.023 \> \,\,0.096 \> -0.875\> \,\,\,\,\,3.768\> \,\,\,-6.313\>
 \,\,\,\,\,3.823\> \,\,0.427\> \,\,1.029\\ 
\> 0.3 \> -0.003 \>  \,\,0.190 \> -2.169\> \,\,\,\,\,9.372\> -15.179\> 
\,\,\,\,\,8.955\> \,\,0.965\> \,\,1.796\\ 
\> 0.5 \> \,\,0.021 \> \,\,0.169 \> -2.675\> \,\,12.596\> -20.429\> \,\,11.943
\> \,\,1.329 \> \,\,2.215\\ 
\> 1.0 \> \,\,0.072 \> \,\,0.261 \> -2.558\> \,\,13.424\> -22.410\> \,\,13.552
\> \,\,1.947\> \,\,2.845\\ 
\> 2.0 \> \,\,0.227 \> \,\,0.404 \> -2.272\> \,\,13.850\> -24.035\> \,\,15.039
\> \,\,2.706\> \,\,3.538\\ 
\> 4.0 \> \,\,0.519 \> \,\,0.523 \> -1.650\> \,\,13.389\> -24.671\> \,\,16.153
\> \,\,3.593\> \,\,4.281\\ 
\> 8.0 \> \,\,0.959 \> \,\,0.498 \> \,\,0.462\> \,\,\,\,\,8.085\> -18.634\>
\,\,14.064\> \,\,4.560\> \,\,5.034\\ 								     
\>\hspace{0.35cm}UV 2800\\						   
\> 0.1 \> -0.015 \> \,\,0.080 \> -0.629\> \,\,\,\,\,2.847\> \,\,\,-5.000\>
\,\,\,\,\,3.131\> \,\,0.374\> \,\,0.931\\
\> 0.3 \> \,\,0.002 \> \,\,0.168 \> -2.048\> \,\,\,\,\,8.843\> -14.342\>
\,\,\,\,\,8.419\> \,\,0.859\> \,\,1.663\\ 
\> 0.5 \> \,\,0.022 \> \,\,0.153 \> -2.733\> \,\,12.578\> -20.239\> \,\,11.694
\> \,\,1.197\> \,\,2.068\\ 
\> 1.0 \> \,\,0.056 \> \,\,0.231 \> -2.705\> \,\,13.627\> -22.465\> \,\,13.406
\> \,\,1.776\> \,\,2.678\\ 
\> 2.0 \> \,\,0.178 \> \,\,0.373 \> -2.466\> \,\,14.133\> -24.140\> \,\,14.899
\> \,\,2.499\> \,\,3.354\\
\> 4.0 \> \,\,0.433 \> \,\,0.512 \> -2.043\> \,\,14.237\> -25.446\> \,\,16.286
\> \,\,3.353\> \,\,4.085\\ 
\> 8.0 \> \,\,0.832 \> \,\,0.517 \> -0.276\> \,\,10.148\> -21.226\> \,\,15.131
\> \,\,4.305\> \,\,4.838\\ 		
\end{tabbing}
\end{table}
\setcounter{table}{4}
\begin{table}
\caption{Coefficients and constants for the calculation of the attenuation of
the thin disk; Continued}
\begin{tabbing}
b \= ssss \= sssssss \= ssssssss \= sssssss \= sssssssss \= ssssssss 
\= ssssssss \= sssssss \= sssssss \kill
\>   \hspace{0.05cm}$\tau^{\rm f}_{\rm B}$
     \>\hspace{0.4cm} $a_0$  \>  \hspace{0.55cm}$a_1$ \> \hspace{0.4cm} $a_2$
     \> \hspace{0.4cm} $a_3$ \>  \hspace{0.65cm}$a_4$ \> \hspace{0.5cm}$a_5$ 
     \> \hspace{0.45cm} $b_0$ \>  \hspace{0.5cm}$b_1$\\ 
\>\hspace{0.35cm}B band\\						     
\> 0.1 \> -0.022 \> \,\,0.116 \> -1.307\> \,\,\,\,\,4.915\> \,\,\,-7.267\>
\,\,\,\,\,3.833\> \,\,0.222\> \,\,0.648\\ 
\> 0.3 \> -0.016 \> \,\,0.215 \> -2.555\> \,\,10.182\> -15.666\> \,\,\,\,\,8.567\>
\,\, 0.571\>  \,\,1.281\\
\> 0.5 \> -0.011 \> \,\,0.257 \> -3.157\> \,\,12.891\> -19.959\> \,\,11.044\>
\,\,0.843\> \,\,1.647\\
\> 1.0 \> \,\,0.009 \> \,\,0.275 \> -3.578\> \,\,15.768\> -24.929\> \,\,14.110
\> \,\,1.320\> \,\,2.207\\
\> 2.0 \> \,\,0.060 \> \,\,0.272 \> -2.665\> \,\,13.784\> -22.905\> \,\,13.793
\> \,\,1.939\> \,\,2.837\\
\> 4.0 \> \,\,0.217 \> \,\,0.411 \> -2.356\> \,\,14.142\> -24.442\> \,\,15.240
\> \,\,2.700\> \,\,3.530\\
\> 8.0 \> \,\,0.509 \> \,\,0.508 \> -1.581\> \,\,13.298\> -24.657\> \,\,16.184
\> \,\,3.587\> \,\,4.272\\						     
\>\hspace{0.35cm}V band\\ 						     
\> 0.1 \> -0.018 \> \,\,0.092 \> -0.954\> \,\,\,\,\,3.339\> \,\,\,-4.755\>
\,\,\,\,\,2.484\> \,\,0.166\> \,\,0.517\\
\> 0.3 \> -0.014 \> \,\,0.168 \> -1.901\> \,\,\,\,\,7.380\> -11.288\> 
\,\,\,\,\,6.205\> \,\,0.444\> \,\,1.086\\ 
\> 0.5 \> -0.013 \> \,\,0.215 \> -2.539\> \,\,10.169\> -15.695\>
 \,\,\,\,\,8.702\> \,\,0.677\> \,\,1.429\\
\> 1.0 \> -0.004 \> \,\,0.246 \> -3.052\> \,\,13.207\> -20.914\> \,\,11.885\>
\,\,1.101\> \,\,1.958\\
\> 2.0 \> \,\,0.027 \> \,\,0.226 \> -2.625\> \,\,13.023\> -21.412\> \,\,12.766
\> \,\,1.656\> \,\,2.559\\ 
\> 4.0 \> \,\,0.129 \> \,\,0.358 \> -2.437\> \,\,13.748\> -23.386\> \,\,14.396
\> \,\,2.356\> \,\,3.226\\
\> 8.0 \> \,\,0.363 \> \,\,0.501 \> -2.060\> \,\,13.957\> -24.816\> \,\,15.836
\> \,\,3.191\> \,\,3.951\\						     
\>\hspace{0.35cm}I band\\ 						   
\> 0.1 \> -0.015 \> \,\,0.099 \> -1.028\> \,\,\,\,\,3.483\> \,\,\,-4.774\>
\,\,\,\,\,2.360\> \,\,0.099\> \,\,0.342\\
\> 0.3 \> -0.013 \> \,\,0.098 \> -1.013\> \,\,\,\,\,4.142\> \,\,\,-6.653\>
 \,\,\,\,\,3.789\> \,\,0.281\> \,\,0.794\\
\> 0.5 \> -0.017 \> \,\,0.161 \> -1.688\> \,\,\,\,\,6.837\> -10.763\>
 \,\,\,\,\,6.030\> \,\,0.451\> \,\,1.092\\
\> 1.0 \> -0.018 \> \,\,0.233 \> -2.399\> \,\,10.019\> -15.881\>
 \,\,\,\,\,9.012\> \,\,0.784\> \,\,1.569\\
\> 2.0 \> -0.010 \> \,\,0.262 \> -2.517\> \,\,11.241\> -18.131\> \,\,10.659\>
\,\,1.245\> \,\,2.121\\
\> 4.0 \> \,\,0.041 \> \,\,0.317 \> -2.517\> \,\,12.639\> -20.955\> \,\,12.677
\> \,\,1.838\> \,\,2.741\\
\> 8.0 \> \,\,0.175 \> \,\,0.424 \> -2.166\> \,\,13.041\> -22.622\> \,\,14.197
\> \,\,2.575\> \,\,3.426\\						     
\>\hspace{0.35cm}J band\\						     
\> 0.1 \> -0.013 \> \,\,0.061 \> -0.619\> \,\,\,\,\,2.041\> \,\,\,-2.712\>
\,\,\,\,\,1.296\> \,\,0.040\> \,\,0.166\\ 
\> 0.3 \> -0.005 \> \,\,0.072 \> -0.811\> \,\,\,\,\,3.134\> \,\,\,-4.751\>
\,\,\,\,\,2.534\> \,\,0.121\> \,\,0.442\\
\> 0.5 \> -0.006 \> \,\,0.092 \> -1.037\> \,\,\,\,\,4.158\> \,\,\,-6.450\>
\,\,\,\,\,3.525\> \,\,0.215\> \,\,0.655\\
\> 1.0 \> -0.007 \> \,\,0.132 \> -1.510\> \,\,\,\,\,6.434\> -10.262\>
\,\,\,\,\,5.739\> \,\,0.420\> \,\,1.036\\
\> 2.0 \> -0.001 \> \,\,0.087 \> -1.678\> \,\,\,\,\,8.377\> -13.945\>
\,\,\,\,\,8.065\> \,\,0.742\> \,\,1.510\\ 
\> 4.0 \> \,\,0.019 \> \,\,0.172 \> -2.137\> \,\,10.787\> -18.017\> \,\,10.624
\> \,\,1.195\> \,\,2.062\\
\> 8.0 \> \,\,0.076 \> \,\,0.266 \> -2.252\> \,\,12.197\> -20.669\> \,\,12.506
\> \,\,1.778\> \,\,2.683\\						     
\>\hspace{0.35cm}K band\\      						     
\> 0.1 \> -0.012 \> \,\,0.071 \> -0.675\> \,\,\,\,\,1.983\> \,\,\,-2.340\> 
\,\,\,\,\,0.982\> \,\,0.004\> \,\,0.051\\ 
\> 0.3 \> -0.001 \> \,\,0.036 \> -0.399\> \,\,\,\,\,1.470\> \,\,\,-2.152\>
\,\,\,\,\,1.094\> \,\,0.018\> \,\,0.158\\
\> 0.5 \> \,\,0.000 \> \,\,0.071 \> -0.753\> \,\,\,\,\,2.703\> \,\,\,-3.856\>
\,\,\,\,\,1.926\> \,\,0.051\> \,\,0.255\\
\> 1.0 \> \,\,0.002 \> \,\,0.112 \> -1.182\> \,\,\,\,\,4.298\> \,\,\,-6.185\>
\,\,\,\,\,3.141\> \,\,0.132\> \,\,0.463\\
\> 2.0 \> \,\,0.007 \> \,\,0.108 \> -1.627\> \,\,\,\,\,6.511\> \,\,\,-9.723\>
\,\,\,\,\,5.091\> \,\,0.278\> \,\,0.777\\
\> 4.0 \> \,\,0.018 \> \,\,0.179 \> -2.214\> \,\,\,\,\,8.920\> -13.524\>
\,\,\,\,\,7.271\> \,\,0.525\> \,\,1.196\\
\> 8.0 \> \,\,0.042 \> \,\,0.254 \> -2.753\> \,\,11.463\> -17.756\>
\,\,\,\,\,9.837\> \,\,0.898\> \,\,1.708\\
\end{tabbing}
\end{table}
\begin{table}[htb]
\caption{Coefficients and constants for the calculation of the attenuation of
the bulge}
\begin{tabbing}
ssssss\= ssss \= ssssssss \= sssssss \= ssssssss \= sssssss \= sssssss \= sssssss 
\= sssssss \kill
\>   \hspace{0.05cm}$\tau^{\rm f}_{\rm B}$
     \>\hspace{0.4cm} $a_0$  \>  \hspace{0.55cm}$a_1$ \> \hspace{0.4cm} $a_2$
     \> \hspace{0.65cm}$a_3$ \>  \hspace{0.7cm}$a_4$ 
     \> \hspace{0.45cm} $b_0$ \>  \hspace{0.5cm}$b_1$\\
\>\hspace{0.35cm}B band\\
\> 0.1 \>	 \,\,0.001 \> \,\,0.045 \> \,\,0.986 \> -2.351 \>  \,\,1.796 \> \,\,0.426 \>  \,\,0.593\\ 
\> 0.3 \>	 \,\,0.029 \> -0.045 \>  \,\,2.003 \> -4.575 \>  \,\,3.662 \>  \,\,0.893 \>  \,\,0.991\\
\> 0.5 \>	 \,\,0.064 \>  \,\,0.079 \>  \,\,1.541 \> -3.613 \>  \,\,3.368 \>  \,\,1.162 \>  \,\,1.187\\
\> 1.0 \>	 \,\,0.176 \>  \,\,0.255 \>  \,\,1.091 \> -1.606 \>  \,\,1.889 \>  \,\,1.516 \>  \,\,1.479\\
\> 2.0 \>	 \,\,0.431 \>  \,\,0.657 \>  \,\,0.859 \> -1.492 \>  \,\,1.670 \>  \,\,1.863 \>  \,\,1.809\\
\> 4.0 \>	 \,\,0.889 \>  \,\,0.580 \>  \,\,1.275 \> -2.861 \>  \,\,2.616 \>  \,\,2.222 \>  \,\,2.181\\
\> 8.0 \>	 \,\,1.333 \>  \,\,0.244 \>  \,\,1.657 \> -3.343 \>  \,\,3.028 \>  \,\,2.616 \>  \,\,2.596\\
\>\hspace{0.35cm}V band\\
\> 0.1 \>	-0.001 \>  \,\,0.059 \>  \,\,0.822 \> -1.943 \>  \,\,1.447 \>  \,\,0.346 \>  \,\,0.497\\	 
\> 0.3 \>	 \,\,0.016 \> -0.019 \>  \,\,1.674 \> -3.820 \>  \,\,3.013 \>  \,\,0.746 \>  \,\,0.881\\
\> 0.5 \>	 \,\,0.040 \> -0.034 \>  \,\,2.106 \> -4.754 \>  \,\,3.876 \>  \,\,1.007 \>  \,\,1.072\\
\> 1.0 \>	 \,\,0.114 \>  \,\,0.270 \>  \,\,0.685 \> -1.475 \>  \,\,2.066 \>  \,\,1.367 \>  \,\,1.350\\
\> 2.0 \>	 \,\,0.296 \>  \,\,0.434 \>  \,\,1.208 \> -1.737 \>  \,\,1.795 \>  \,\,1.714 \>  \,\,1.662\\
\> 4.0 \>	 \,\,0.674 \>  \,\,0.651 \>  \,\,1.225 \> -2.523 \>  \,\,2.308 \>  \,\,2.065 \>  \,\,2.014\\
\> 8.0 \>	 \,\,1.153 \>  \,\,0.416 \>  \,\,1.316 \> -2.906 \>  \,\,2.756 \>  \,\,2.440 \>  \,\,2.409\\
\>\hspace{0.35cm}I band\\
\> 0.1 \>	-0.002 \>  \,\,0.081 \>  \,\,0.598 \> -1.408 \>  \,\,1.001 \>  \,\,0.246 \>  \,\,0.361\\
\> 0.3 \>	 \,\,0.007 \>  \,\,0.003 \>  \,\,1.318 \> -2.947 \>  \,\,2.225 \>  \,\,0.533 \>  \,\,0.699\\
\> 0.5 \>	 \,\,0.018 \> -0.022 \>  \,\,1.708 \> -3.796 \>  \,\,2.966 \>  \,\,0.755 \>  \,\,0.886\\
\> 1.0 \>	 \,\,0.056 \>  \,\,0.016 \>  \,\,2.029 \> -4.501 \>  \,\,3.771 \>  \,\,1.114 \>  \,\,1.148\\
\> 2.0 \>	 \,\,0.153 \>  \,\,0.404 \>  \,\,0.308 \> -0.590 \>  \,\,1.486 \>  \,\,1.468 \>  \,\,1.433\\
\> 4.0 \>	 \,\,0.387 \>  \,\,0.517 \>  \,\,1.041 \> -1.362 \>  \,\,1.488 \>  \,\,1.812 \>  \,\,1.754\\
\> 8.0 \>	 \,\,0.808 \>  \,\,0.628 \>  \,\,1.127 \> -2.431 \>  \,\,2.306 \>  \,\,2.166 \>  \,\,2.116\\
\>\hspace{0.35cm}J band\\
\> 0.1 \>	-0.003 \>  \,\,0.129 \>  \,\,0.299 \> -0.799 \>  \,\,0.543 \>  \,\,0.156 \>  \,\,0.218\\	   
\> 0.3 \>	 \,\,0.001 \>  \,\,0.094 \>  \,\,0.692 \> -1.659 \>  \,\,1.214 \>  \,\,0.311 \>  \,\,0.446\\
\> 0.5 \>	 \,\,0.007 \>  \,\,0.070 \>  \,\,1.003 \> -2.342 \>  \,\,1.764 \>  \,\,0.449 \>  \,\,0.609\\
\> 1.0 \>	 \,\,0.025 \>  \,\,0.044 \>  \,\,1.505 \> -3.440 \>  \,\,2.704 \>  \,\,0.728 \>  \,\,0.862\\
\> 2.0 \>	 \,\,0.072 \>  \,\,0.097 \>  \,\,1.715 \> -3.873 \>  \,\,3.292 \>  \,\,1.087 \>  \,\,1.125\\
\> 4.0 \>	 \,\,0.182 \>  \,\,0.352 \>  \,\,0.925 \> -2.026 \>  \,\,2.323 \>  \,\,1.447 \>  \,\,1.406\\
\> 8.0 \>	 \,\,0.414 \>  \,\,0.663 \>  \,\,0.047 \>  \,\,0.198 \>  \,\,0.727 \>  \,\,1.790 \>  \,\,1.721\\
\>\hspace{0.35cm}K band\\
\> 0.1 \>	-0.003 \>  \,\,0.146 \>  \,\,0.132 \> -0.435 \>  \,\,0.267 \>  \,\,0.099 \>  \,\,0.121\\
\> 0.3 \>	 \,\,0.002 \>  \,\,0.135 \>  \,\,0.256 \> -0.712 \>  \,\,0.486 \>  \,\,0.154 \>  \,\,0.213\\
\> 0.5 \>	 \,\,0.006 \>  \,\,0.125 \>  \,\,0.373 \> -0.973 \>  \,\,0.693 \>  \,\,0.208 \>  \,\,0.296\\
\> 1.0 \>	 \,\,0.016 \>  \,\,0.105 \>  \,\,0.637 \> -1.561 \>  \,\,1.166 \>  \,\,0.333 \>  \,\,0.468\\
\> 2.0 \>	 \,\,0.037 \>  \,\,0.082 \>  \,\,1.040 \> -2.468 \>  \,\,1.920 \>  \,\,0.549 \>  \,\,0.703\\
\> 4.0 \>	 \,\,0.081 \>  \,\,0.095 \>  \,\,1.418 \> -3.340 \>  \,\,2.754 \>  \,\,0.871 \>  \,\,0.963\\
\> 8.0 \>	 \,\,0.171 \>  \,\,0.258 \>  \,\,1.095 \> -2.666 \>  \,\,2.632 \>  \,\,1.242 \>  \,\,1.229\\
\end{tabbing}